# ReLink: Computational Circular Design of Planar Linkage Mechanisms Using Available Standard Parts

**First author**
Maxime Escande*, Email: *mescande@ethz.ch*, ORCID: *https://orcid.org/0000-0002-1385-3083*

**Last author**
Kristina Shea, Email: *kshea@eth.ch*, ORCID: *https://orcid.org/0000-0003-3921-2214*

**Affiliation**
Engineering Design and Computing Laboratory, Department of Mechanical and Process Engineering, ETH Zurich, 8092 Zurich, Switzerland

* Corresponding author.

## Abstract

The Circular Economy framework emphasizes sustainability by reducing resource consumption and waste through the reuse of components and materials. This paper presents ReLink, a computational framework for the circular design of planar linkage mechanisms using available standard parts. Unlike most mechanism design methods, which assume the ability to create custom parts and infinite part availability, ReLink prioritizes the reuse of discrete, standardized components, thus minimizing the need for new parts. The framework consists of two main components: design generation, where a generative design algorithm generates mechanisms from an inventory of available parts, and inverse design, which uses optimization methods to identify designs that match a user-defined trajectory curve. The paper also examines the trade-offs between kinematic performance and $CO_2$ footprint when incorporating new parts. Challenges such as the combinatorial nature of the design problem and the enforcement of valid solutions are addressed. By combining sustainability principles with kinematic synthesis, ReLink lays the groundwork for further research into computational circular design to support the development of systems that integrate reused components into mechanical products.

**Keyword**: Circular economy – Design for circularity – Computational circular design – Reuse – Computational kinematics – Linkages

## 1. Introduction

The rising pressures on the environment and the depletion of natural resources (Rockström et al., 2009) have increased the urgency for sustainable solutions. Among these, the Circular Economy (CE) has emerged as a promising framework to reduce resource consumption, pollution, and waste by promoting closed material loops (Ellen MacArthur Foundation, 2013; Swiss Federal Council, 2021). However, current public initiatives have emphasized mainly low-value recovery strategies, such as recycling processes that, while mitigating waste, often capture a fraction of the material's original value. This is especially true for high-performance materials like carbon-fiber reinforced polymer (CFRP) and aluminum alloy, which are increasingly used in lightweight structural components due to their high strength-to-weight ratio. For example, linkages made from CFRP and aluminum alloys have been employed in lightweight robotic arms, where reducing inertial loads is critical (Wang et al., 2025; Yin et al., 2019). Due to the irreversible cross-linking of thermosetting resins, recycling CFRP parts typically results in significant degradation of the fibers, relegating them to downcycled, low-value applications rather than enabling high-end reuse (Isa et al., 2022). Such limitations underscore the need for higher-value retention strategies, such as remanufacturing and repurposing, that preserve the intrinsic value of a product's parts by allowing them to be effectively reintegrated into new products.

Remanufacturing and repurposing both focus on reusing components of existing products to create new ones with the same and different functionalities, respectively.

A promising showcase is the reuse of building components in the construction sector (Brütting et al., 2019b; Zhang and Shea, 2024), where the static nature of the components allows reuse to focus primarily on geometric compatibility. However, mechanical products, such as those requiring moving components, must be designed at a system level and currently lack systematic methodologies to support circular design. Linkage mechanisms, which are essential in industries such as automotive, agriculture, and manufacturing, are traditionally designed with the assumption that part geometries and quantities can be tailored to meet the design requirements, rather than being constrained by availability (Deshpande and Purwar, 2020; Huang and Campbell, 2015; Lumpe and Shea, 2023; Nobari et al., 2024, 2022; Nurizada et al., 2025). In practice, however, creating new parts incurs both economic costs and significant environmental impacts.

To address this gap, this work introduces ReLink, a computational approach to the synthesis of planar linkage mechanisms using available parts from an inventory. In contrast to the assumption of unlimited part availability, ReLink follows a constrained design paradigm that prioritizes the reuse of discrete, standardized parts over the creation of new parts. This approach is illustrated in the circular lifecycle shown Figure 1. At the end of a mechanism's lifecycle, disassembly and part property acquisition close the loop, thus preserving both the value and the embodied impacts of the components. This paper focuses on the upper part of the cycle proposed, starting from an inventory of available parts, their assembly combinations, and the synthesis of linkage mechanisms.

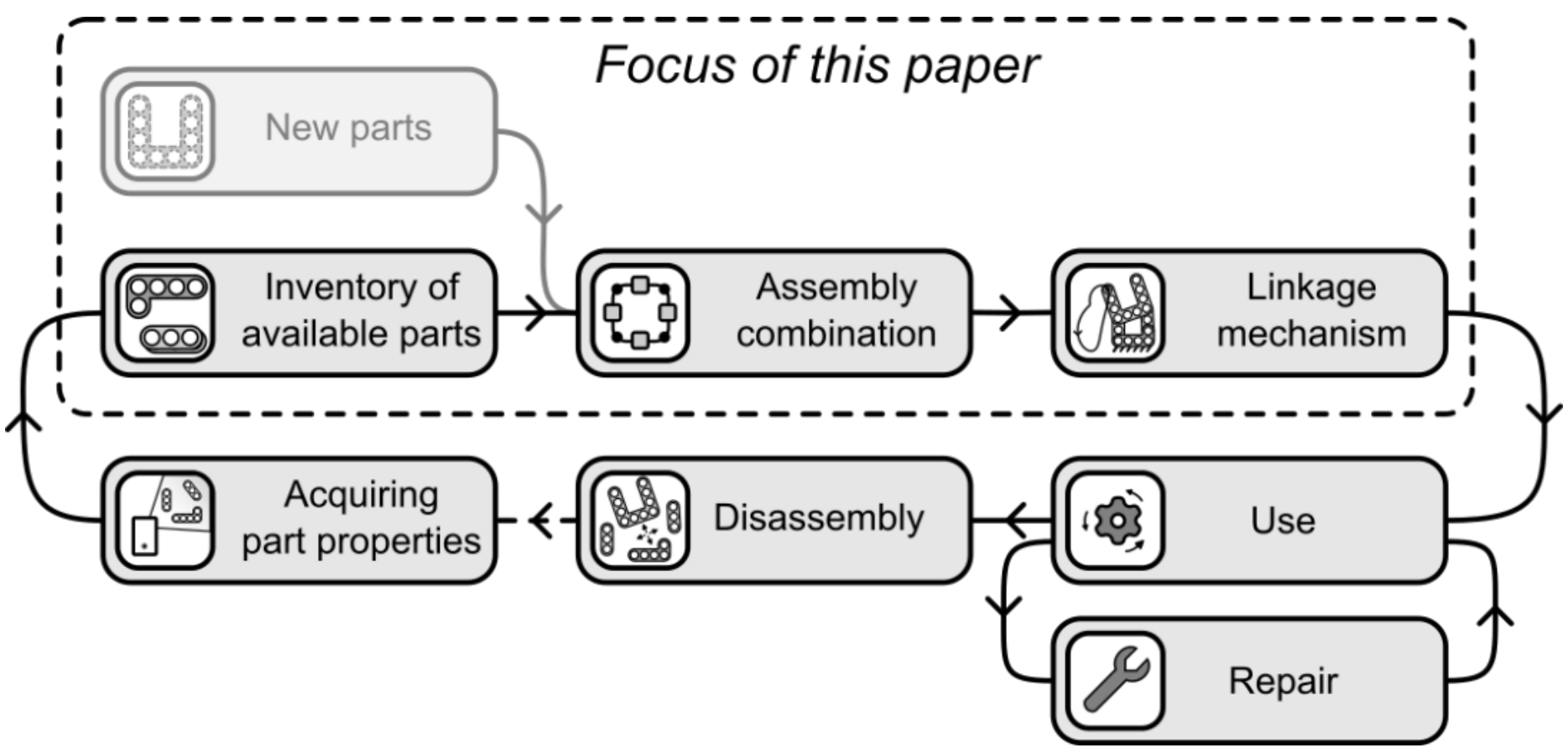

*Figure 1: Illustration of the ReLink framework with a circular design scenario for the reuse of standard parts. The paper focuses on the upper part of the cycle proposed.*

The presented framework consists of two key elements: design generation and inverse design. An overview is presented in Figure 2. The available part set consists of LEGO® Technic beams[1]. Using these parts allows the ReLink framework to be demonstrated for a limited, well-defined set of parts, while also serving as a symbolic representation of reusable components. During the design generation process, a generative algorithm is employed to automatically create linkage mechanisms using parts from the available inventory. The trajectories of the part of the mechanisms are evaluated by a kinematic solver. In the inverse design, an optimization framework is employed to find mechanisms that closely approximate a user-defined curve. The inverse design problem addresses two distinct scenarios: one constrained to using only available parts in the inventory and another that incorporates new parts while considering their environmental impact. Both are modeled as integer, non-linear, and non-differentiable minimization problems.

[1] LEGO® is a trademark of the LEGO Group, which does not sponsor or endorse this work. The content and interpretations resented in this paper are solely those of the authors and do not reflect the views or opinions of the LEGO Group.

The use of LEGO® Technic beams in the model is not only for illustration. These modular components share key characteristics with standardized parts found in many industries. For instance, many mechanical systems incorporate components like beams or engine linkage elements that come in fixed sizes and connection configurations. By showing that a limited, well-defined inventory can be assembled into complex linkage mechanisms, ReLink offers a practical demonstration of how high-value components can be repurposed in new applications. This work, therefore, not only addresses the design challenge for mechanisms but also provides insights that can be transferred to broader sustainable manufacturing practices, potentially enabling circular reuse strategies in sectors that rely on high-performance materials.

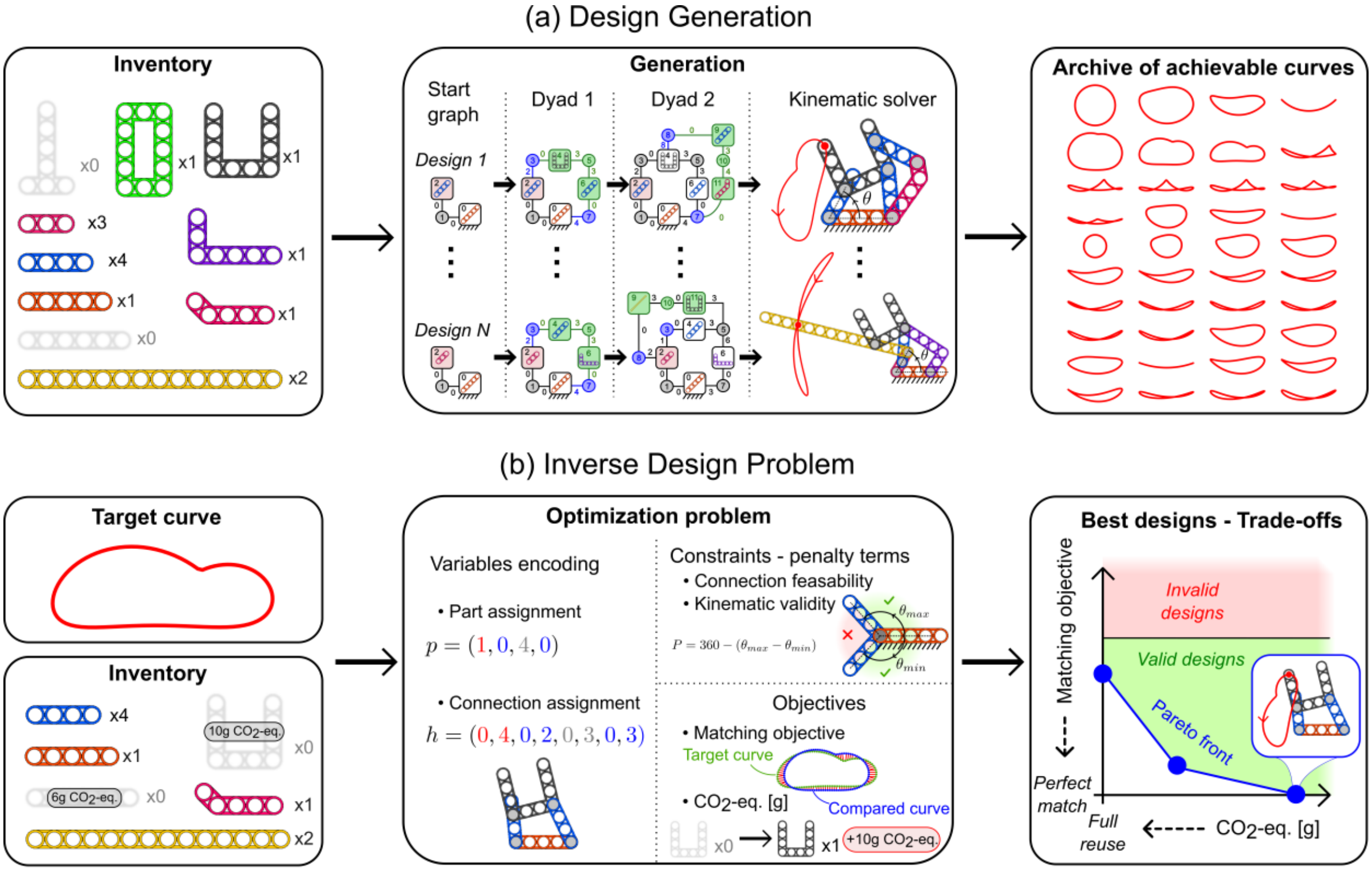

*Figure 2: Overview of the two problems solved in this paper. (a) Mechanisms are generated from an existing part inventory of LEGO technics parts. A kinematic solver evaluates the curves achievable with the mechanisms. An archive of mechanisms and curves is created. (b) The inverse design method requires a target curve and a part inventory as inputs. The optimization then finds a set of Pareto optimal designs based on maximizing shape matching and minimizing CO2.*

### 1.1. Contributions

By bridging design for circularity and kinematics, this paper defines a new benchmark problem for component reuse in product design. The key contributions are twofold. First, from a sustainability and circularity perspective:

- The mechanisms generated mainly involve **available parts**, adhering to the principle of "form follows availability" (Gorgolewski, 2017). This supports resource conservation and part reuse, and results in a new dataset for computational circular design.

- The paper examines **tradeoffs** between the performance and impact of designs when combining new and available parts.

Second, from the mechanism perspective, contrary to most kinematic generators and solvers in the literature:

- The parts are **multimodal**, meaning they can be connected at different locations, not just at the ends. This allows for more complex assemblies and is closer to industrial scenarios where parts can connect at multiple points.
- The parts can take on **diverse shapes**, extending beyond simple straight bars. Parts can be shaped as “I” (straight), “L”, “J”, “T”, “U”, and “O” (rectangle).
- The parts are **discrete** elements, as opposed to bars with continuously variable lengths. Their geometry and connection points are constrained to specific values (e.g., straight bars of lengths 2,3,4, ...). This makes the problem fully combinatorial and non-differentiable, thus significantly increasing its complexity.

Many industrial systems rely on standardized mechanical components like connecting rods, struts, and brackets, whose dimensions and connection modalities are predetermined for manufacturing consistency and performance. In this sense, the constraints imposed with LEGO® Technic mirror those of actual, off-the-shelf parts.

# 2. Background and Related Works

## 2.1. Circular Economy and Design for Circularity

Potting et al. (2017) present a list of nine “R” circular strategies from the highest to the lowest value retention. Initially, public efforts have been focused on lower value retention strategies like R9 Recover (incineration of waste) and R8 Recycle. However, higher value retention solutions are needed to better close material flows and reduce the need for energy. Some strategies rely more on a change of habit and consumption patterns (e.g.: R0 Refuse, R2 Reduce, R3 Reuse), while some others require more technical and design considerations.

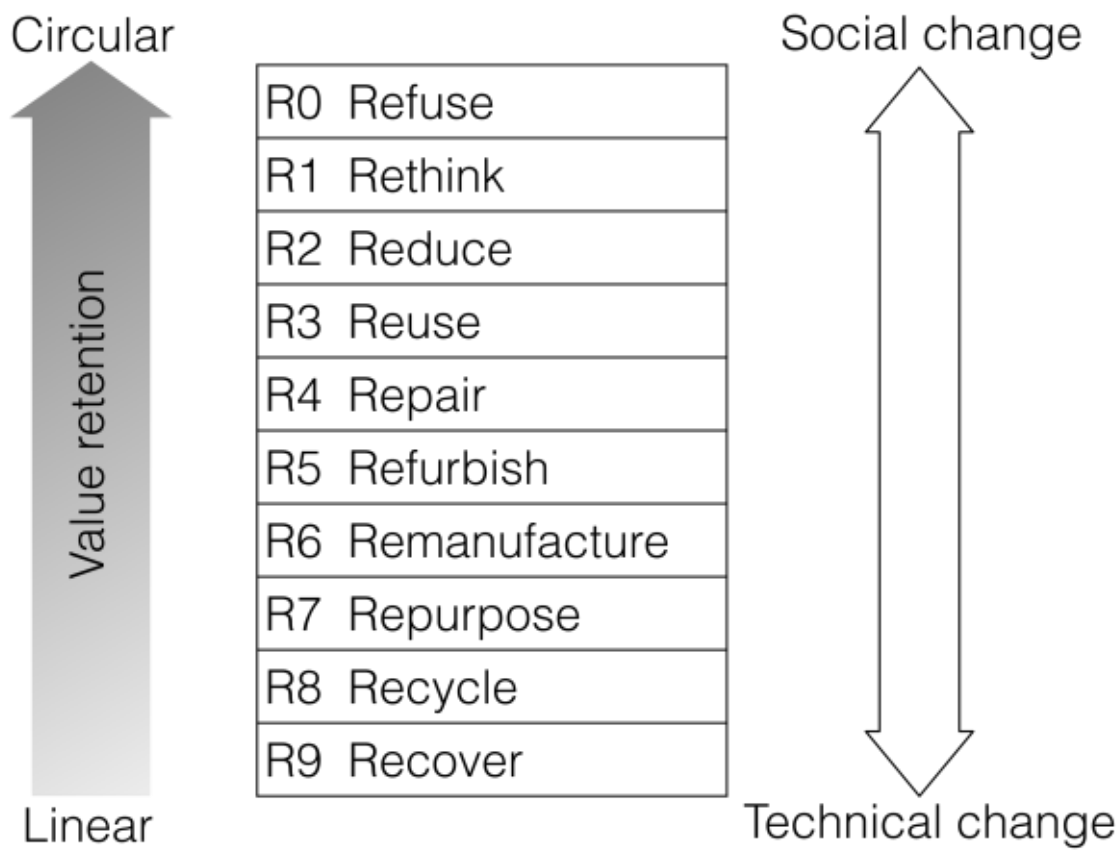

*Figure 3: R- strategies for circular economy. The left arrow highlights that the first strategies have higher value retention. The right arrow indicates the general direction of change needed to achieve a strategy. Adapted from Potting et al. (2017)*

Since most of the impact of a product is determined at the early design phases (Tischner, 2017), circularity should be integrated in the design practices for new products (Mesa, 2023). However, the circular economy is not only about creating better new products but also about extending the use of existing resources. A circular approach can also be applied to extend the lifespan of existing products (e.g.: R4 Repair, R5 Refurbish) (Richter et al., 2023; van Oudheusden et al., 2023), and produce new life cycles. For instance, R6 (Remanufacture) and R7 (Repurpose) involve reusing existing parts of a product to create another product with the same function or a different one, respectively. Unlike Reuse where the entire product is directly reused as-is, e.g., secondhand use, Remanufacture and Repurpose involve recombining components from disassembled products to create new ones. For clarity, this paper refers to Repurpose and Remanufacture collectively as “part reuse” or “component reuse”, distinguishing them from R3.

The approach of reusing parts is already being explored in the construction sector. For example, Raghu et al. (2022) proposed a machine learning method to automatically extract building characteristics from Google Street View, an approach referred to as “existing building as material bank”, which is particularly relevant for component reuse. Brütting et al. (2019a) and Huang et al. (2021) present methods for the reuse of parts for new structures as an optimization problem. In a design process where “form follows availability”, Gorgolewski (2017) inverts the traditional process—where a prescribed design is created first and resources are then extracted and manufactured accordingly—into a design process where available materials and components inform and shape the design from the start, encouraging creativity within the constraints of part reuse. Brütting et al. (2020) compare the Life Cycle Assessments (LCA) of load-bearing structures in different scenarios including using only new parts, using only reused parts and a mix of both. Similarly, Devos et al. (2024) perform an LCA to compare between reclaimed and new ceramic bricks.

## 2.2. Product design

Within the mechanical design community, the concept of circular economy is also being explored. Ortner et al. (2022) present a computational method for circular product design with a Multi-Objective Optimization (MOO). Their case study focuses on the modular design of furniture, demonstrating how MOO can navigate and balance multiple, potentially conflicting or redundant circular economy objectives to improve design outcomes. Bakker et al. (2014) compare different scenarios between prolonging the lifespan of a device or upgrading it to benefit from efficiency improvements.

In mechanical product design, some projects highlight the reuse of parts through remanufacturing and repurposing. For instance, Bridgens et al. (2018) list several repurposed products as a way to promote better resource use and empower communities through making. In industry, KYBURZ Switzerland AG, a manufacturer of high-quality electric vehicles, has introduced its “MultiLife” concept, an approach aimed at refurbishing vehicles and components for extended use (KYBURZ, 2024).

Several computational frameworks have been proposed in the scientific literature to integrate remanufacturing considerations into early-stage product design. Behtash et al. (2023) introduce a remanufacturing co-design framework that simultaneously addresses product design and remanufacturing decisions. Building on this, Liu et al. (2025) present a structured Design for Remanufacturing (DfRem) process, organized into sequential stages of identification, evaluation, and validation stages. Other studies focus on specific computational challenges within remanufacturing workflows: Kim et al. (2021) develop a unified approach that leverages customer feedback from online reviews to guide the configuration of remanufactured products, while also determining optimal disassembly strategies and levels for harvesting components from end-of-life products. Complementing this, Xia et al. (2020) propose an optimization method that matches reused parts with similar remaining lifespans to extend product life and make better use of available components. Together, these works highlight diverse yet complementary strategies for systematically incorporating remanufacturing considerations into mechanical design.

Despite these efforts, the mechanical design community still lacks systematic methods for identifying new uses for repurposed parts, more specifically to recombine parts in new products. Hewa Witharanage et al. (2024) identify the lack of methods to support idea generation using repurposed parts as a major research gap. The present work aims to address this gap by introducing a part reuse framework where LEGO® Technic beams are used to represent standardized parts to be reused.

A review of relevant works for LEGO® automated design and LEGO® reuse highlights some promising steps in this direction. One example is Brickit, a mobile app that detects hundreds of LEGO® bricks from a photo and suggests templates and instructions for designs that can be built with the available bricks (Brickit, 2024). However, these design templates are mostly static and must be curated in advance. Alternatively, Xu et al. (2019) present a computational method for creating LEGO®

Technic models based on a 3D wireframe sketch. While this method maps designs to a predefined brick set, it does not constrain the number of available parts. Building on these ideas, this work introduces a more constrained approach to exploring part reuse in product design considering availability.

### 2.3. Mechanism design and optimization

In the mechanical engineering design community, automated design has evolved significantly, leveraging computational methods to generate and optimize mechanical systems. Pioneering works in computational design synthesis employ algorithmic approaches to explore vast design spaces efficiently (Cagan et al., 2005; Campbell et al., 2003; Chakrabarti et al., 2011). Building on these foundations, Shea and colleagues introduced graph grammar-based techniques that enable the systematic creation of complex mechanical structures through formalized rule sets for a wide range of examples including bicycle frames, gearboxes and soft robots (Königseder and Shea, 2014; van Diepen and Shea, 2019).

Within the broader field of automated mechanical design, mechanism design represents a well-established and actively studied subdomain, particularly the four-bar linkage, one of the most thoroughly investigated mechanisms (McCarthy, 2016). A four-bar linkage consists of four rigid links connected in a loop by four joints, typically configured to allow rotational movement. This configuration transforms input motion into a desired output motion, making it essential for applications in automobiles, machinery, and robotics (Me Virtuoso, 2023). The study of mechanisms involves two core aspects: kinematic analysis and kinematic synthesis.

Kinematic analysis involves evaluating a given mechanism by determining the motion of each member within it. Typically, a mechanism is driven by a rotating actuator as input and the output can be defined by a coupler curve, which is a trajectory described by a point on a linkage. Two types of solvers can be used for kinematic analysis. Direct, analytical solvers, like the dyadic decomposition, which is also called arc-intersection, propagate the positions and orientations of known nodes to adjacent dyads geometrically (Bächer et al., 2015). A dyad is a set of two parts connected together. Iterative solvers, on the other hand, rely on optimization techniques to handle more complex assemblies where a dyadic decomposition is not possible. For example, Lyu et al. (2023) developed a real-time kinematic simulator for planar mechanisms that incorporates both methods, while Nobari et al. (2022) generated a dataset of one hundred million planar mechanisms and their corresponding kinematics using the dyadic decomposition approach. This paper employs the dyadic decomposition method due to its robustness, speed, and the wide diversity of mechanisms that can be decomposed into dyads (Nobari et al., 2024).

Kinematic synthesis is an inverse design process, where a target kinematics, such as a coupler curve, is provided as input, and the objective is to design a mechanism whose output closely matches the input. For mechanisms with bars of continuous length, the problem can be formulated as a differentiable optimization problem (Bächer et al., 2015). Other approaches leverage machine learning; for instance, Deshpande and Purwar (2020) used a variational auto-encoder (VAE) to create a low-dimensional encoding of coupler curves, facilitating their retrieval from a database. Similarly, Nobari et al. (2024) employed contrastive learning to jointly map mechanisms and coupler curves, enabling the retrieval of multiple candidate mechanisms for an input curve, which were subsequently refined through optimization.

Despite notable efforts to integrate sustainability and kinematics, the body of research remains limited. For example, He et al. (2020) propose a lightweight, underactuated robotic ankle exoskeleton optimized for a reduced carbon footprint. To the best of the authors' knowledge, no methods have been developed for the circular design of mechanisms that specifically focus on reusing parts, supporting sustainability through the repurposing of existing components.

### 2.4. Challenge

The challenge of finding optimal designs given a fixed set of parts requires managing the **combinatorial complexity** of the problem. As the number of parts in both the inventory and the design increases, the number of possible combinations grows exponentially. This complexity is further compounded by the possibility of connecting parts at various locations, rather than just at the ends. A simple example illustrates this challenge:

Consider an inventory of $N_t = 5$ different available part types and no limitation on quantity. Let each design consist of $N_p = 4$ parts (for instance, a four-bar linkage). Each part type has $N_l = 15$ possible connection locations, and $N_c = 2$ other parts connected to each location. The number of possible part assignments for a design is given by ${N_t}^{N_p} = 5^4 = 625$. The number of possible connections for a design is $\left({N_l}^{N_c}\right)^{N_p} = (15^2)^4 \approx 2.5 \cdot 10^9$. Even with this simple example, the total number of combinations exceeds 1.5 trillion. Using a vectorized and parallelized solver on a GPU taking around $20\mu s$ per mechanism (Nobari et al., 2022), evaluating the kinematics of all combinations would take roughly a year. As such, a naive enumeration approach is not realistic.

Another challenge arises from the non-linear nature of the objective function. For example, in the design of truss structures with a discrete set of available parts, a reformulation could lead to a Mixed Linear Integer Program (MILP) (Brütting et al., 2018; Rasmussen and Stolpe, 2008), which is efficient to solve using a branch-and-cut solver. However, for mechanisms, the kinematics are non-linear and sometimes even non-differentiable (Bächer et al., 2015), making the optimization problem more complex.

The final challenge is the intricate structure of the design space. Among all possible designs, only a fraction are valid (Nobari et al., 2022). Even among the valid ones, variations can lead to invalid configurations in a highly complex and discontinuous manner (Nobari et al., 2024).

## 3. Methods

The method section follows the framework as shown in Figure 2. The design generation, including design representation, graph generation, kinematic solver, curve matching procedure, and dataset generation, is presented in Sections 3.1 to 3.5. The inverse design problem, using only available parts or allowing trade-offs by combining new and existing parts, is presented in Section 3.6.

### 3.1. Design Representation

The mechanisms consist of parts and pin connectors. The parts are a collection of LEGO® Technic beams, as illustrated in Figure 4. The parts are connected at their holes by pin connectors, essentially constraining two or more holes to each other and keeping the rotational degree of freedom.

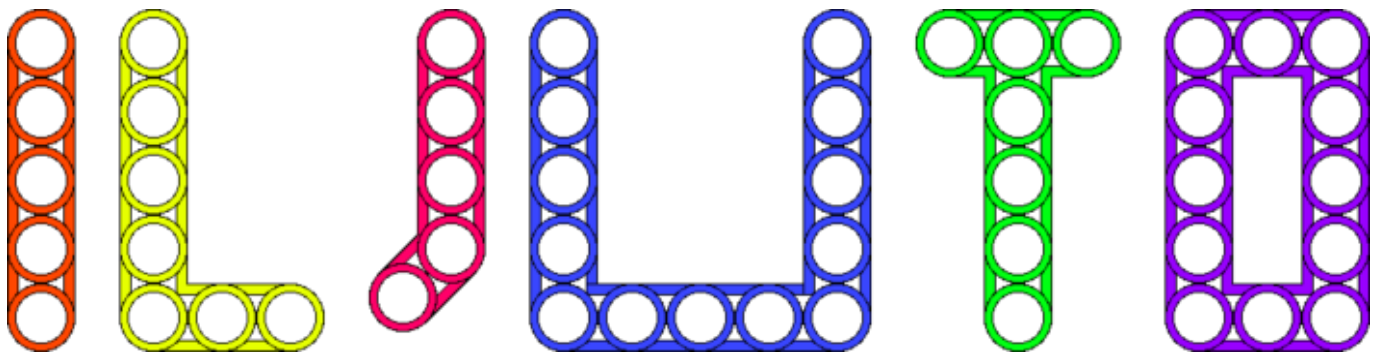

*Figure 4: Different types of parts modeled in this paper, shown from left to right: "I", "L", "J", "U", "T", and "O" types. Specific lengths and geometries can vary.*

The mechanisms are represented as graphs. Since parts can be connected to multiple pin connectors and a pin connector can connect multiple parts, both parts and pin connectors are represented as vertices. The edges in the undirected simple graph signify the physical connections between a part and a pin connector. The edge attribute indicates the hole index in the part to which the pin connector is connected. To fully determine the mechanism, one part is defined to be the ground, and one part is defined to be the actuator. The actuator part has an additional attribute $\theta$, representing the angle between actuator and ground. Figure 5 shows both the graph representation of a connection between two parts through a pin connector, and an assembly with six parts.

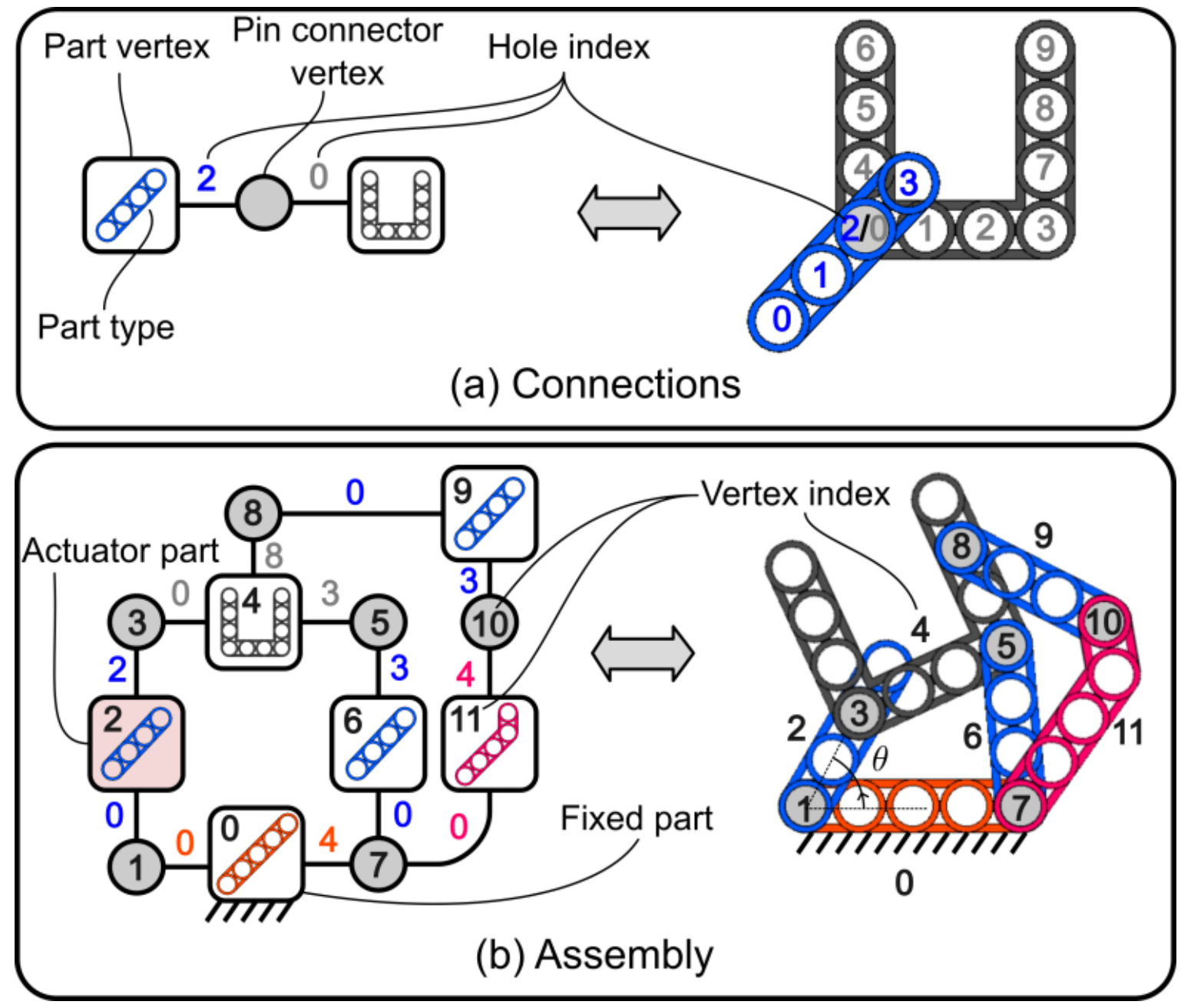

*Figure 5: (a): Example graph of two parts connected by a pin connector. (b): Example graph of a mechanism with six parts and six pin connectors. The mechanisms contain a ground part and an actuator part with defined angle θ to the ground part. The indices in parts and pin connectors vertices are the vertex index, not to be confused with the hole indices shown near the edges.*

### 3.2. Graph generation

The graphs are generated procedurally and randomly by following the rule "add dyad", similar to Nobari et al. (2022). The process begins with a start graph that includes the fixed part and the actuator. A dyad consists of two parts connected by a pin connector. To incorporate a dyad to an existing graph, two parent pin connectors are either added to existing parts or selected among the existing pin connectors. The child vertices forming the dyad are then connected to the parent pin connectors. Figure 6 illustrates the process of generating a mechanism with six parts through two dyad additions.

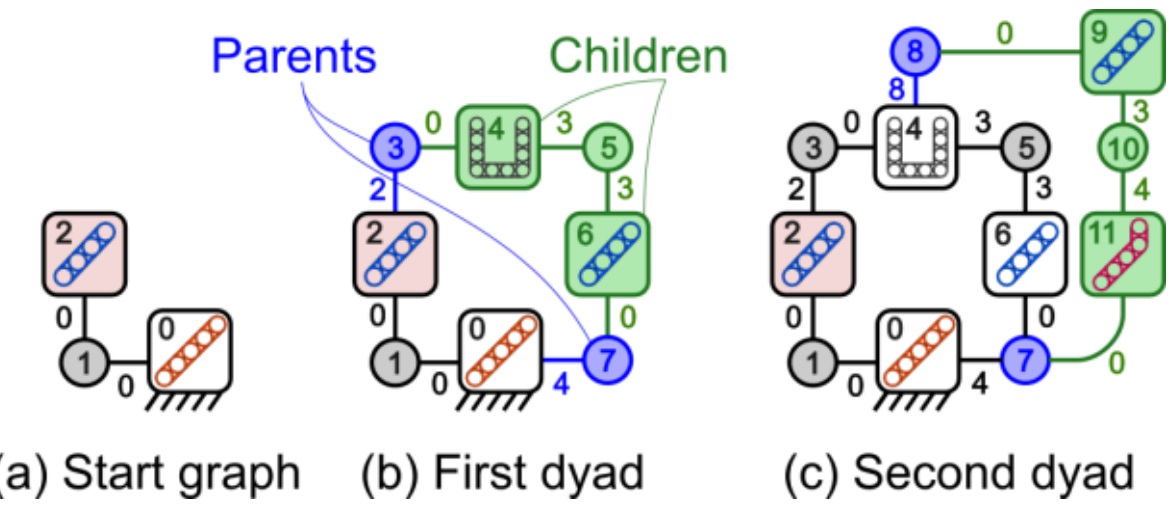

*Figure 6: Generating a (b) four-bar linkage and (c) six-bar linkage.*

By applying this rule, the topology of a four-bar linkage is achieved after one dyad addition to the start graph, while a six-bar linkage is obtained after a second dyad addition.

### 3.3. Kinematic Solver

The kinematic analysis of a mechanism involves determining the positions and orientations of all parts for all actuator angles. Due to the graph construction based on dyads, the arc intersection method (Bächer et al., 2015) can be directly applied in the same order as the dyads were added. This dyadic construction not only simplifies the solver but also guarantees that the generated mechanisms have a single degree of freedom, corresponding to the rotation of the actuator part.

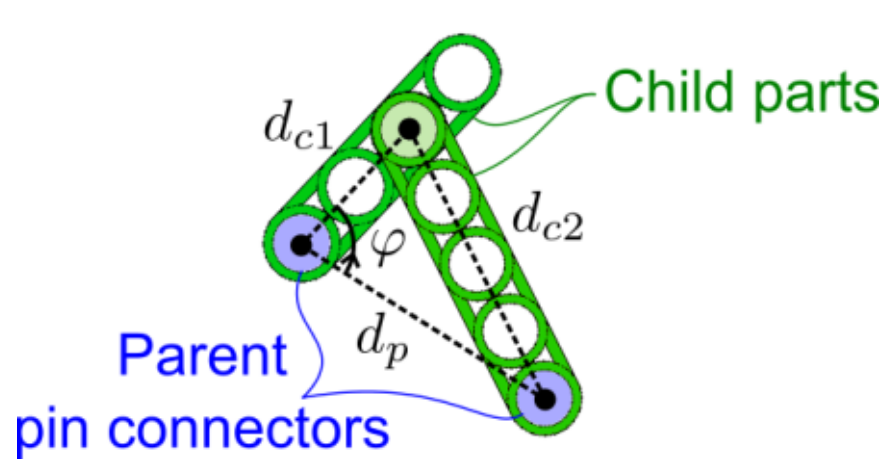

*Figure 7: Illustration of the use of the arc intersection to determine the orientation of a dyad. The parent pin connectors from previously solved dyads or from the start graph uniquely determine the position of the child parts.*

Figure 7 demonstrates the kinematic solving process for a single dyad. The positions of the parent pin connectors are either known from a previously solved dyad or provided by the ground and actuator part of the start graph. The positions of the child parts are determined since the connecting hole must align with parent pin connectors' positions. The orientation of the child parts is deduced from the Law of cosines:

$$cos\,\varphi = \frac{d_p^2 + d_{c1}^2 - d_{c2}^2}{2 d_p d_{c1}} \tag{1}$$

Here, $d_p$ is the distance between the parent pin connectors, $d_{c1}$ and $d_{c2}$ are the distances between the selected holes on children 1 and 2, respectively, and $\varphi$ is the angle between the parent pin connectors and the selected holes on child 1.

A dyad can be solved if and only if the following condition is met:

$$|cos\,\varphi| < 1 \tag{2}$$

The solving condition (2) allows the detection of invalid states in the mechanism. An invalid state $|\cos\varphi| \geq 1$ occurs when the dyad cannot close because its elements are either too close or too far apart to connect. The operating range $[\theta_{min}, \theta_{max}]$ of a mechanism is defined as the longest continuous interval of actuator angles during which no invalid state is encountered.

### 3.4. Normalization and Matching

Within a mechanism's operating range, trajectories are derived by tracking the positions of the holes in the parts (Figure 8(a)). Only the trajectories of parts with complex motions are further considered, specifically simple rotations and fixed parts are excluded.

The trajectories are normalized through a multi-step process, which normalizes the rotation and position but preserves trajectory scale. Scale preservation is essential since the parts are not scale independent. The trajectory global position and orientation are arbitrary, allowing mechanisms to be compared independently of their placement or rotation. The normalization procedure consists of four steps: *trajectory resampling*, *position normalization*, *orientation normalization* and *direction alignment* illustrated in Figure 8(b), and detailed in the Supplementary Information (SI) Section A.1.

Once normalized, the similarity between the curves is measured using the Chamfer Distance (CD) according to Equation 3, where a lower CD corresponds to a higher similarity (Figure 8(c)).

$$\mathrm{CD}(A, B) = \frac{1}{n_A} \sum_{p_A \in A} \min_{p_B \in B} d(p_A, p_B) + \frac{1}{n_B} \sum_{p_B \in B} \min_{p_A \in A} d(p_A, p_B) \tag{3}$$

where $p_A \in A$ are the $n_A$ points of curve $A$, $p_B \in B$ are the $n_B$ points of curve $B$, and $d(p_A, p_B)$ is the Euclidean distance between points $p_A$ and $p_B$.

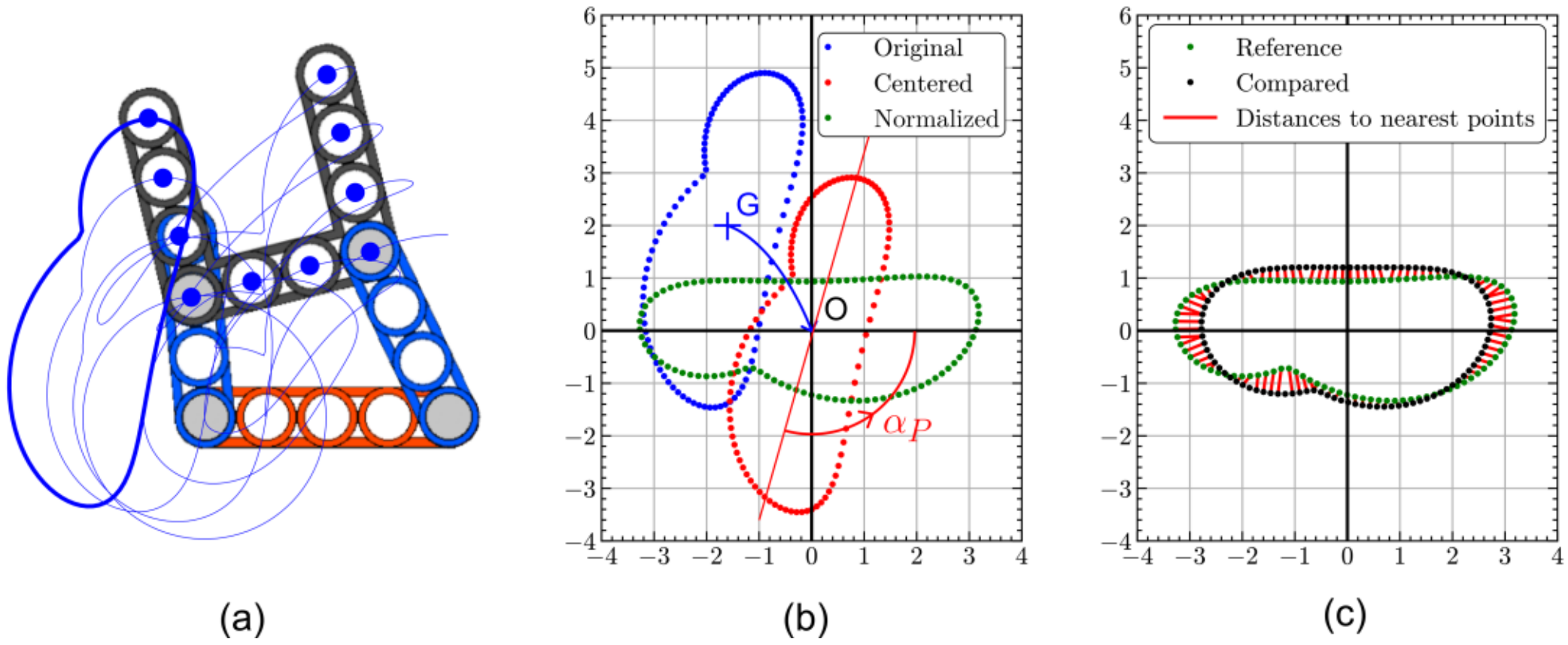

*Figure 8: (a) Evaluation of the trajectories of a mechanism. (b) Normalization of one curve by moving it to the origin and aligning rotation and direction (c) Curve similarity expressed by the Chamfer Distance.*

### 3.5. Dataset Generation

The dataset for kinematic curves generated by mechanisms from parts in a defined inventory is illustrated in Figure 2(a). As mentioned in Section 2.4, a full enumeration of all possible mechanisms is not realistic. Instead, an archive of a random subset of mechanisms is generated. The algorithm for the archive creation takes as input the inventory set, the number of mechanisms to generate, the number of dyads in the mechanisms, and the number of data points per curve. The algorithm randomly assembles mechanisms from available parts, evaluates all non-trivial coupler curves, normalizes and saves them and corresponding mechanisms. This procedure is presented in Figure 9.

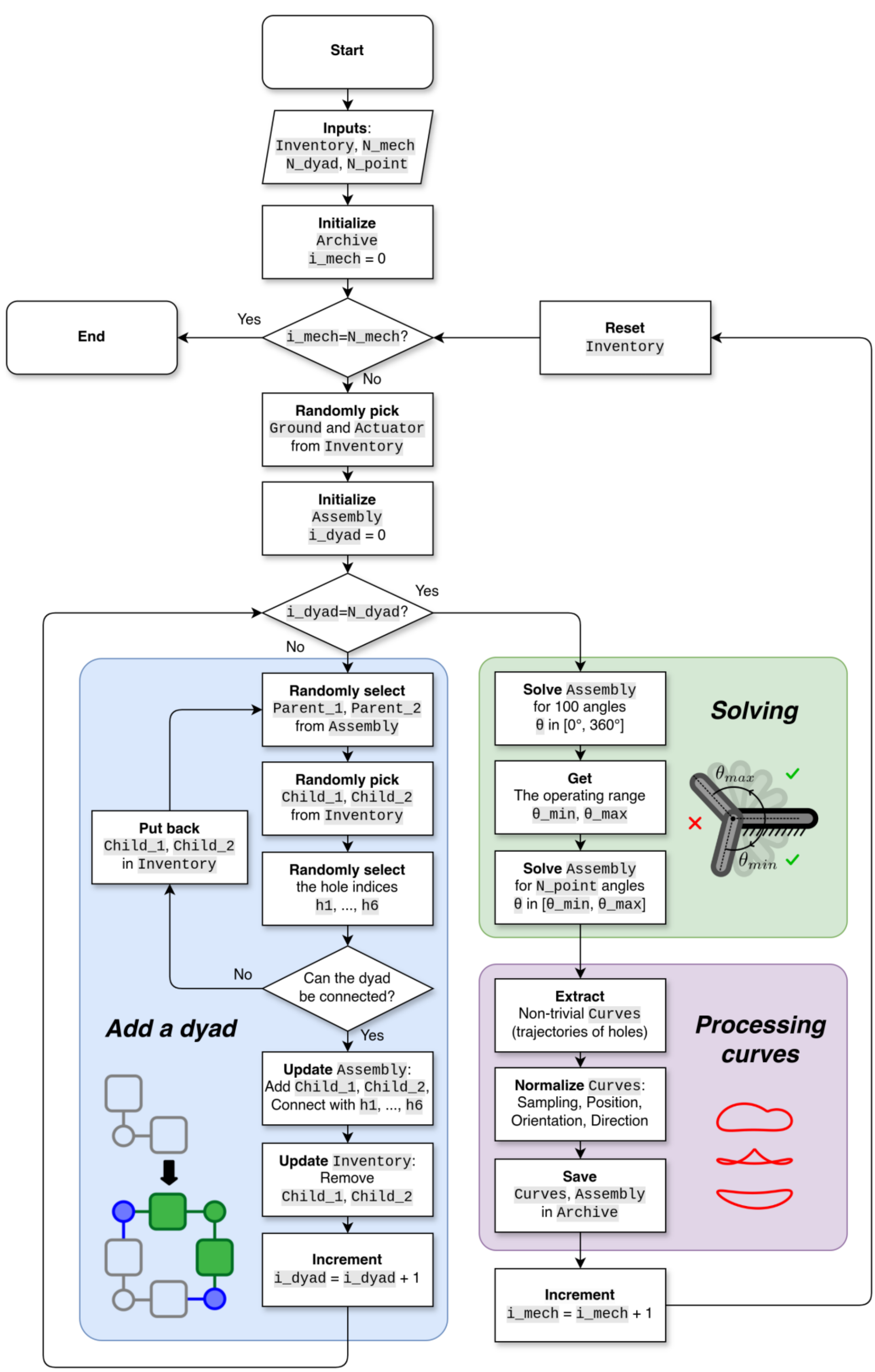

*Figure 9: Procedure for generating mechanism designs and related normalized couple curves from an inventory.*

### 3.6. Inverse Design Setting

The inverse design problem aims to find the mechanism composed of parts from the inventory to produce a curve similar to a target curve (see Figure 2(b)). This is tackled with two different approaches. In the first approach, only available parts from the inventory are allowed. The second approach explores trade-offs with both available and new parts allowed. In this paper, both inverse design approaches are demonstrated using four-bar linkage topologies, i.e. a single dyad. However, the inverse design approach is presented in a generalized way to hold for any topology.

The inverse design is formulated as an optimization scheme. This requires a formulation of the mechanisms suitable for optimization. The topology of a mechanism is represented by its graph $\mathcal{G}$, with $V$ part vertices, $V'$ pin connector vertices, and $E$ edges, where each edge connects a part and a pin connector. The degree of the vertex $i$ is noted $d_i$. The number of different part types is noted $N$.

For optimization, the parts are reformulated as a vector of integers $\boldsymbol{p} \in \{0, \dots, N-1\}^V$ indicating which part types $p_i$ from the inventory are assigned to the $i$-th part vertex of the graph. The holes are reformulated analogously as a vector of integers $\boldsymbol{h} \in \{0, \dots, C-1\}^E$, where $C$ is the maximum hole count and $h_i^{j \in \{0,\dots,d_i-1\}}$ is the index of the hole represented by the $j$-th edge of the $i$-th part vertex. $d_i$ is the degree of this part vertex. The pin connectors are not further considered since they have no influence on the kinematics. In the case of four-bar linkages, the encoded mechanisms have 12 integer variables $(V = 4, E = 8)$. As an example, Figure 10 shows the mechanism encoded by $\boldsymbol{p} = (1,0,4,0)$, $\boldsymbol{h} = (0,4,0,2,0,3,0,3)$

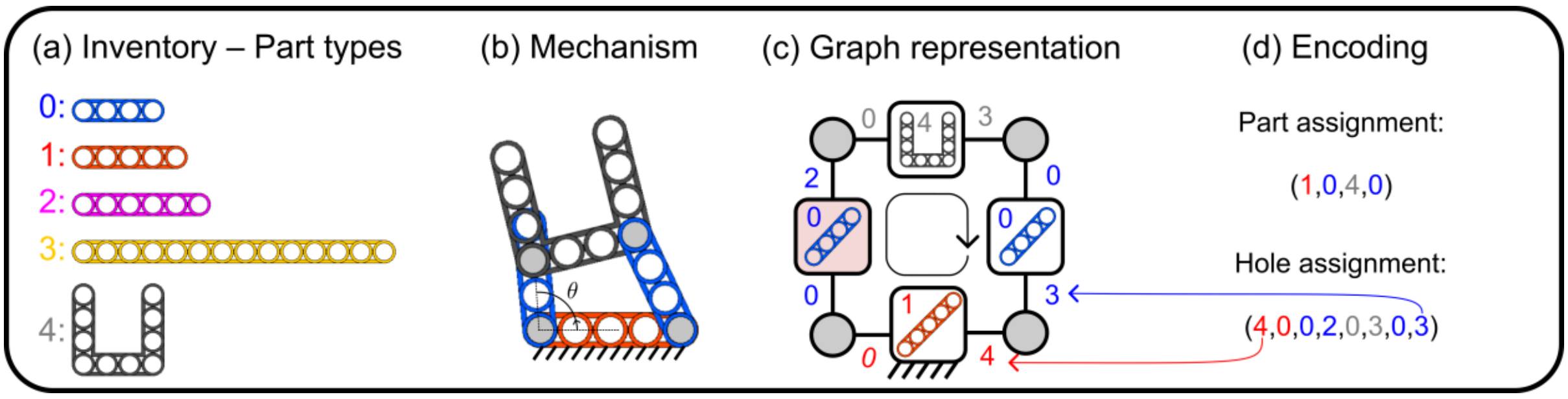

*Figure 10: For a given inventory (a), the linkage mechanism shown in (b), and represented by a graph (c) is encoded with the part and hole assignment vectors (d).*

### 3.7. Single-Objective Optimization: Curve Matching with Available Parts

The objective of the first optimization problem is to identify a mechanism with a coupler curve closely matching the target curve, while strictly adhering to the available parts in the inventory. The objective function is thus formulated as a weighted sum of the curve matching and penalty terms according to Equation 4, where $w_{\mathrm{CD}}$ and $w_1$ to $w_4$ are the weights associated to each term, $f_{\mathrm{CD}}$ is a term associated to the curve matching and $P_1$ to $P_4$ are penalty terms.

$$\min_{\boldsymbol{p},\boldsymbol{h}} f_{\mathrm{kin}}(\boldsymbol{p}, \boldsymbol{h}) = w_{\mathrm{CD}} \cdot f_{\mathrm{CD}} + w_1 P_1 + w_2 P_2 + w_3 P_3 + w_4 P_4 \tag{4}$$

The curve matching term $f_{\mathrm{CD}}$ is expressed according to Equation 5, which evaluates all coupler curves $C_i$ generated from $\boldsymbol{p}$ and $\boldsymbol{h}$ and normalizes the Chamfer Distance between zero and one using the hyperbolic tangent function.

$$f_{\mathrm{CD}}(\boldsymbol{p}, \boldsymbol{h}) = \tanh\left(\min_{C_i \in \mathcal{C}(\boldsymbol{p},\boldsymbol{h})} \mathrm{CD}(T, C_i)\right) \tag{5}$$

The penalty terms $P_1$ to $P_4$ are penalizing the infeasible mechanisms. $P_1$ counts the number of parts in the current mechanism, which are not in the inventory. $P_2$ penalizes double assignment of holes in a part to multiple pin connectors. $P_3$ penalizes assignment of pin connectors to non-existent holes and $P_4$ penalizes if a mechanism does not have a 360° of operating range. The weight and the penalty terms considered in this problem are listed in SI Section A.2.

In the remainder of the paper, designs respecting the part availability, hole availability and hole existence conditions ($P_1 = P_2 = P_3 = 0$) are described as "admissible". Admissibility is a necessary condition for a design to be valid, but not a sufficient one. The mechanism may still become locked at certain angles or even be geometrically impossible to assemble. If the design is not admissible, the geometric validity penalty and the curve matching term are directly set to their respective maximum values $P_4 = 360$ and $f_{CD} = 1$. This approach ensures that only admissible designs are solved. Contrary to the generation method presented in Section 3.5 where only admissible designs are generated with the "add dyad" rule, the inverse design requires the three penalty terms $P_1$, $P_2$, and $P_3$ to be evaluated to verify that a given combination of parameters $(\boldsymbol{p}, \boldsymbol{h})$ is admissible.

### 3.8. Multi-Objective Optimization: Exploring Design Trade-Offs

The second optimization problem allows the inclusion of unavailable parts in the design. In this formulation, the original parts availability penalty $P_1$ is transformed into an additional objective, which minimizes the greenhouse gas (GHG) footprint associated with creating new parts. The GHG impact of LEGO® Technic beams is evaluated from the LCA results of a study comparing the environmental impact of toys (Levesque et al., 2022). The study reports a GHG footprint of 3.2 g $CO_2$ equivalent per gram of ABS plastic used in LEGO® toys. The mass of a LEGO® Technic beam is estimated to be proportional to the number of connection holes, with a mass of 0.26g/hole (BrickLink, 2001). This yields a GHG impact of a new part $p_{new}$, measured in grams of $CO_2$ equivalent, that can be expressed according to Equation 6, where $c_{new}$ is the number of holes of the new part.

$$\text{GHG}(p_{new}) = c_{new} \cdot 0.83 \tag{6}$$

The **GHG objective** function $f_{\text{GHG}}(\boldsymbol{p})$ depends on the part assignment vector $\boldsymbol{p}$. It follows a structure similar to $P_1(\boldsymbol{p})$, as it penalizes parts used beyond the available inventory. Specifically, the GHG impact of parts exceeding the available quantity is given by:

$$f_{\text{GHG}}(\boldsymbol{p}) = \sum_{\text{count}(k) > n_k} \text{GHG}(\boldsymbol{p}_k) \cdot (\text{count}(k) - n_k) \tag{7}$$

where $\text{count}(k)$ is the number of part type $k$ used in the design and $n_k$ denotes the quantity of part type $k$ available in the inventory. The part type $k$ exists in the inventory even if it is unavailable. This means that all part types to be considered by the algorithm must be included in the inventory, even if $n_k = 0$.

The resulting problem is formulated as a **multi-objective optimization** with two objectives:

$$\begin{gathered}\min_{\boldsymbol{p},\boldsymbol{h}} \hat{f}_{\text{kin}}(\boldsymbol{p}, \boldsymbol{h}) = w_{\text{CD}} \cdot f_{\text{CD}} + w_2 P_2 + w_3 P_3 + w_4 P_4 \\ \min_{\boldsymbol{p}} f_{\text{GHG}}(\boldsymbol{p})\end{gathered} \tag{8}$$

The only difference between $f_{\text{kin}}$ and $\hat{f}_{\text{kin}}$ is the absence of the part availability penalty term $P_1$ in $\hat{f}_{\text{kin}}$. For simplicity, both kinematic objectives are collectively referred to as $f_{\text{kin}}$ in the remainder of this paper.

### 3.9. Optimization Strategies

The two inverse design problems, single-objective (4) and multi-objective (8), are both black-box, non-linear minimization problems with integer variables. The single-objective problem is solved using the three solving strategies detailed below:

- **Random Search**: This is a breadth-first search with no depth. Designs are randomly generated by first filling $\boldsymbol{p}$ with random available part types, and second by randomly selecting the terms in $\boldsymbol{h}$ such that the number of available holes in the related part is respected. This grants that the designs are admissible: $f_{\text{kin}}(\boldsymbol{p}, \boldsymbol{h}) \leq 560$. The objective functions are evaluated for each random design, progressively populating the design space.

- **Random Greedy Search**: This is a depth-first search with as much breadth as authorized by the maximum evaluation. An admissible design is randomly generated and evaluated. Its direct neighbors are then evaluated, and the best-performing (based on $f_{\text{kin}}$) neighbor is selected. This process continues iteratively, with each new design's neighbors being evaluated until no further improvement is possible. The procedure is repeated using new random starting designs. The direct neighbors of a design are defined as the designs obtained by varying one variable.

- **Genetic Algorithm (GA)**: A population of designs evolves through simulated mutations and crossovers, following the principle of "survival of the fittest", using the python library PyGAD (Gad, 2024). The parameters selected after a parameter study are presented in SI Section A.3.

The multi-objective problem is solved with a genetic algorithm only, also using PyGAD, but with NSGA-II selection strategy (Deb et al., 2002).

# 4. Results and Discussion

The results of this study are divided into four parts. Section 4.1 details the generation of archives of achievable curves, following the method introduced in Section 3.5. Section 4.2 evaluates and compares the performance of the different solving strategies presented in Section 3.9. Section 4.3 explores trade-offs between kinematic performance and GHG impact. Finally, Section 4.4 discusses challenges encountered during the process, including issues related to enumeration, redundancy, sensitivity, and scarcity.

## 4.1. Archive of Achievable Curves

The procedure outlined in Figure 9 is applied to generate archives of mechanisms and their corresponding coupler curves, using only parts from the inventory shown in Figure 11.

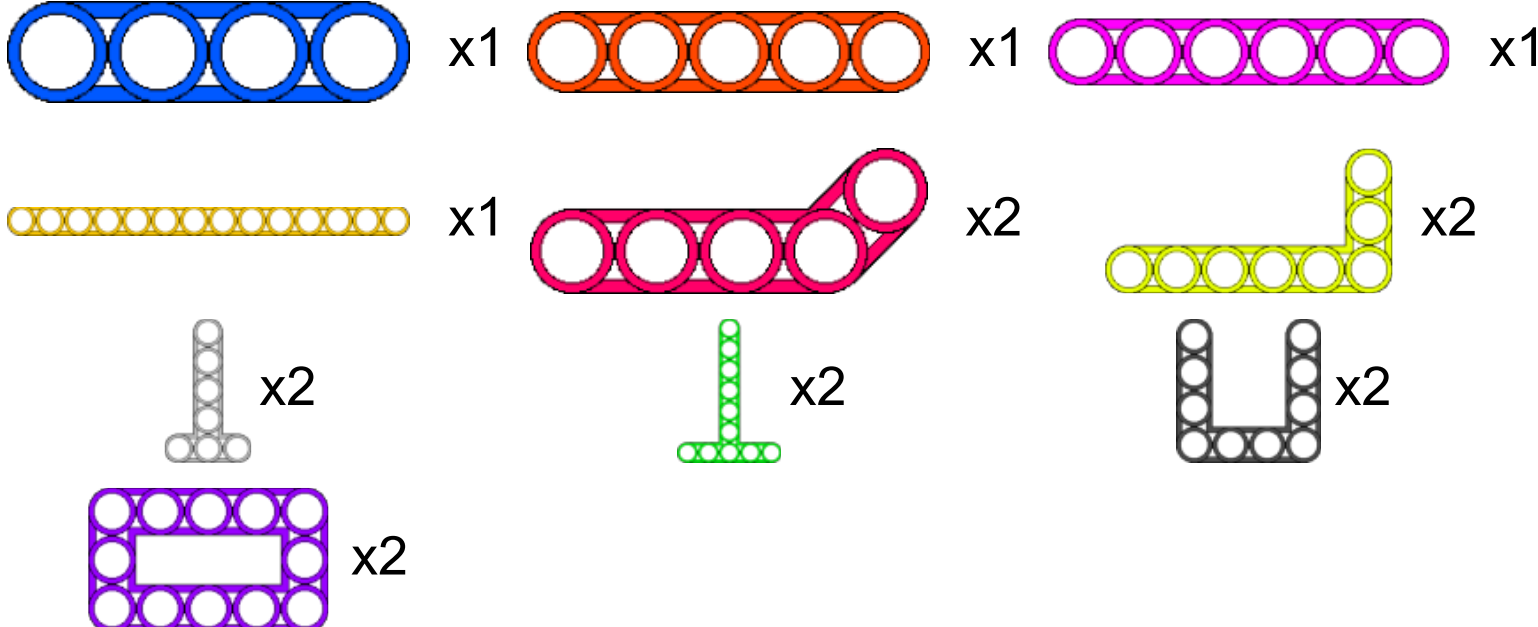

*Figure 11: Inventory of part types used for the generation of the archives. The parts are not shown to scale relative to each other.*

With the addition of two dyads (six parts in the assembly), 1000 mechanisms are generated, producing a total of 12535 coupler curves in 19 seconds on a Dell XPS 13 9310 laptop equipped with an 11th Gen Intel® Core™ i7-1195G7 processor (single-core performance). A random selection of 600 curves from this archive is illustrated in Figure 12.

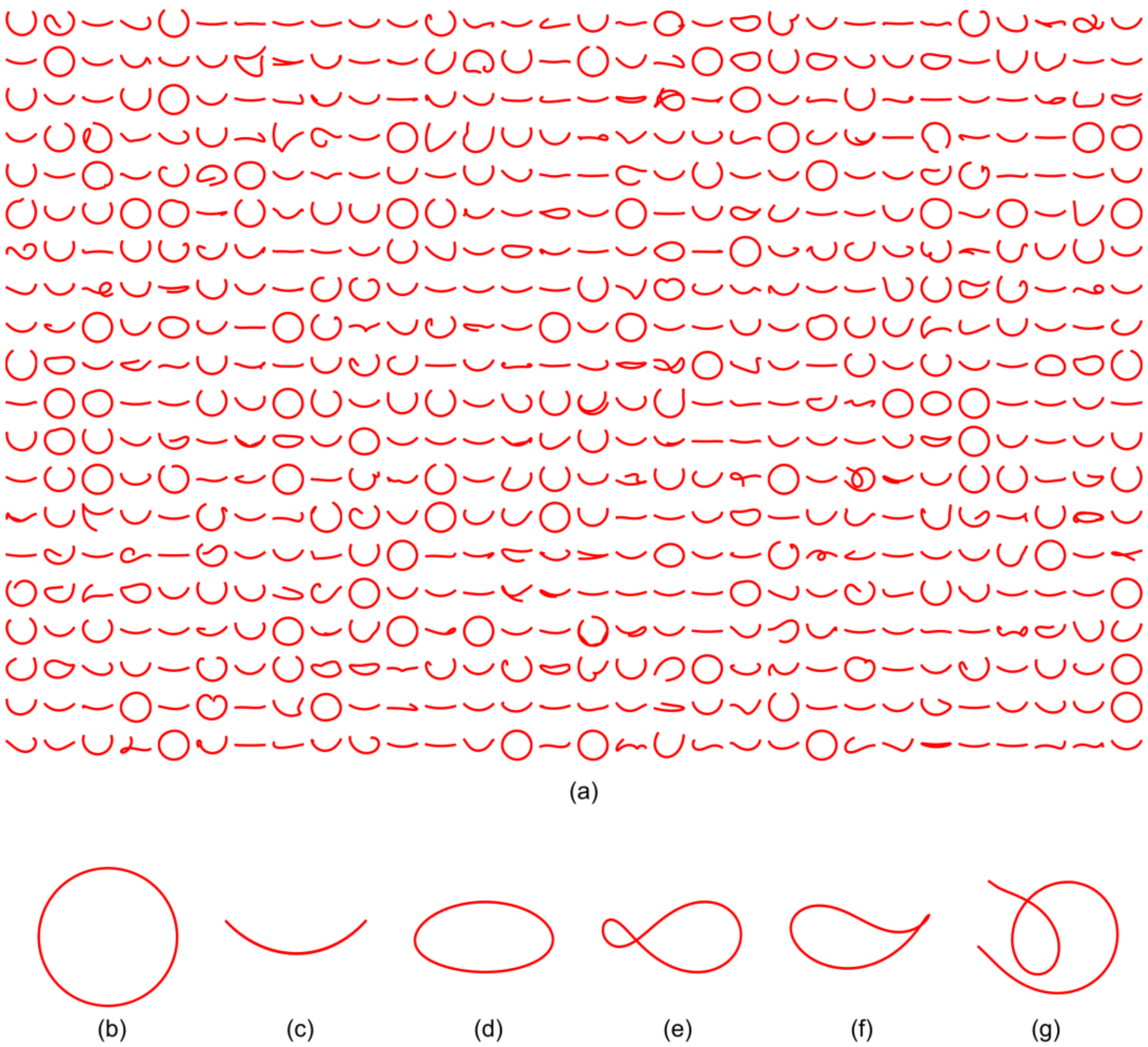

*Figure 12: (a) A random selection of coupler curves from mechanisms generated with two dyads. For clarity, the curves are not shown to scale relative to each other. (b-g) show representative examples selected from the set in (a).*

The extract of the archive shown in the Figure 12 reveals a diversity of coupler curves, including both closed and not closed curves. These include simple geometric shapes such as circles (b) and circular arcs (c), as well as more complex forms like pseudo-ellipses (d), lemniscates (e), airfoil-like shapes (f), and intricate, irregular curves (g). This range highlights the richness and versatility of mechanisms composed of two dyads, with the given inventory, capable of producing a broad spectrum of coupler curve geometries.

To quantify these results, Table 1 compares the statistics of archives generated with 1, 2 and 5 dyads, each consisting of 1000 randomly generated mechanisms.

*Table 1: Statistics of the curve archives for different numbers of dyads. The best values are in bold.*

| **Number of dyads added** | 1 | 2 | 5 |
|---|---|---|---|
| **Average number of curves per mechanism** | 8.5 | 12.5 | **19.6** |
| **Mean operating range** | **232°** | 203° | 163° |
| **Percentage of closed curves** | **27.8%** | 19.3% | 11.5% |
| **Percentage of curves that are part of a circle** | **30.5%** | 34.8% | 43.1% |

Several trends can be observed from Table 1. Mechanisms with more dyads tend to generate a higher number of couple curves, on average, as additional dyads increase the number of hole positions to track. However, the addition of more dyads also raises the likelihood of mechanism lock-up, which reduces the mean operating range. Consequently, the percentage of closed curves decreases with more dyads.

Having too many open curves may be undesirable when the goal is to actuate the mechanism with a motor, as locked mechanisms are less practical for continuous operation. Similarly, the increasing percentage of curves that are part of a circle, full circles or arcs, may reduce the archive's diversity, which could be a drawback depending on the design objectives.

Nevertheless, the archive generated with two dyads still contains over 2400 closed curves and more than 8100 curves that are not part of a circle. Given that the archive was generated in just a few seconds on a standard laptop, these results are both satisfactory and highlight the efficiency of the proposed method in producing feasible designs and coupler curves for a specified inventory.

### 4.2. Comparing Solving Strategies

To benchmark the three solving strategies for the single-objective inverse design problem presented in Section 3.7, two specific closed curves are selected as targets, and a reduced inventory is defined, as shown in Figure 13. The chosen topology is a four-bar linkage. Both target curves are achievable with the given inventory but require a precise selection of part types and connections.

Target curve (a) is chosen for its larger scale, as its mechanism requires at least two of the longest parts. Target curve (b) is selected for its specificity, as its mechanism requires a "U"-shaped part as a coupler link.

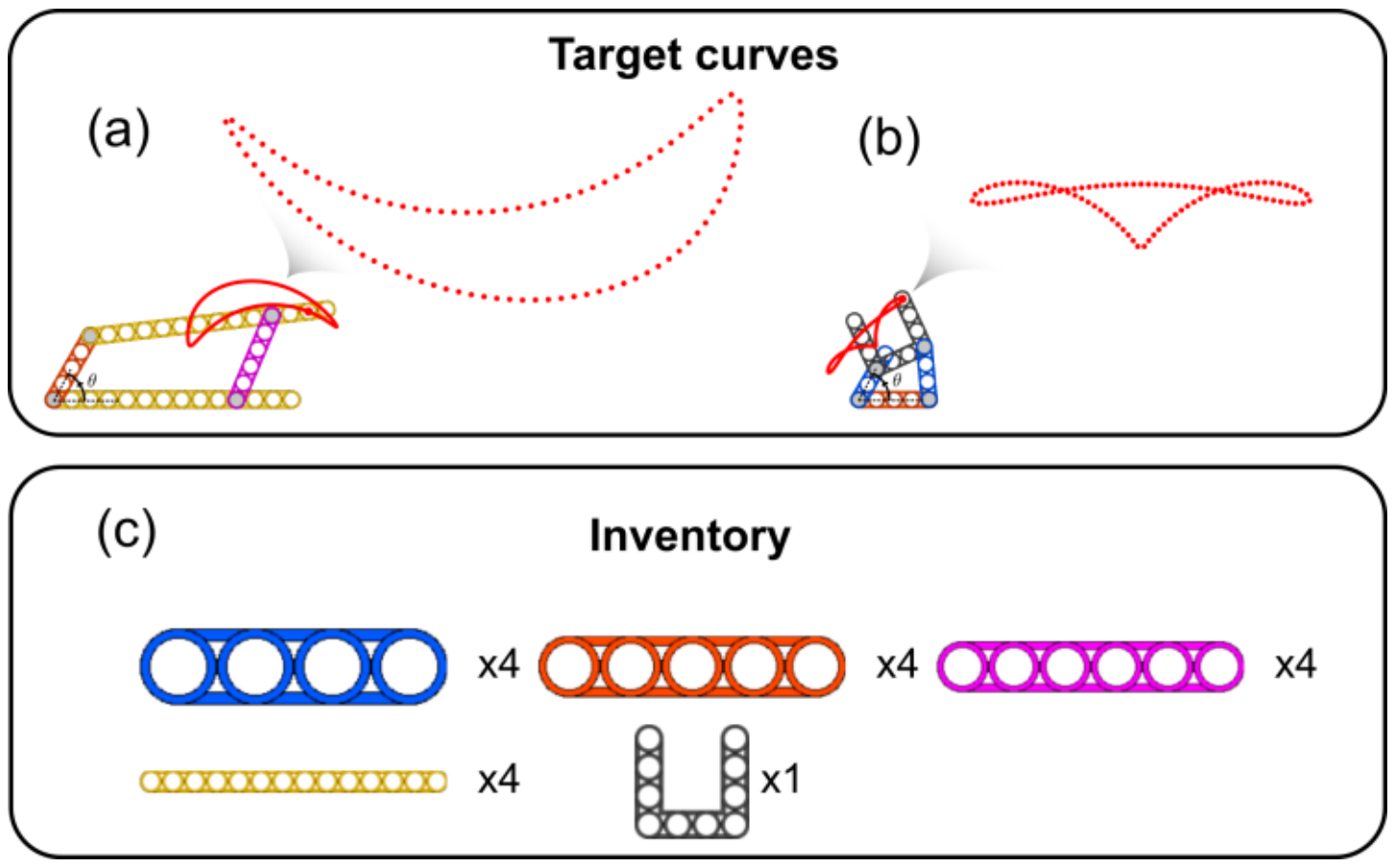

*Figure 13: Overview of the inverse design task for the single-objective optimization. The upper part illustrates the target curves: (a) has an airfoil-like shape, while (b) includes two crossings and a cusp. The relative scale between the two curves is preserved. Part (c) shows the reduced inventory.*

For both target curves, the three solving strategies (random search, random greedy search and genetic algorithm) are each allocated $N_{\text{eval}} = 10{,}000$ objective function evaluations to find the best matching design, based on the single-objective formulation in Equation (4). Each experiment is repeated 100 times to ensure repeatability and to determine median and interquartile ranges for the convergence behavior of each method. SI Section A.3 summarizes the parameters used with the genetic algorithm.

Figure 14 shows the results of the three solving strategies for the two target curves. To compare the convergence of the three solvers, (a.1) and (b.1) display the evolution of the best objective values encountered across the design evaluations. The medians and interquartile ranges are used to evaluate variability across repeated optimizations, as they are less influenced by outliers and directly relate to designs encountered. Specifically, median designs, that is those associated with the median final objective values, represent designs where 50% of the final solutions are equal or more performant, and 50% are less performant within each series.

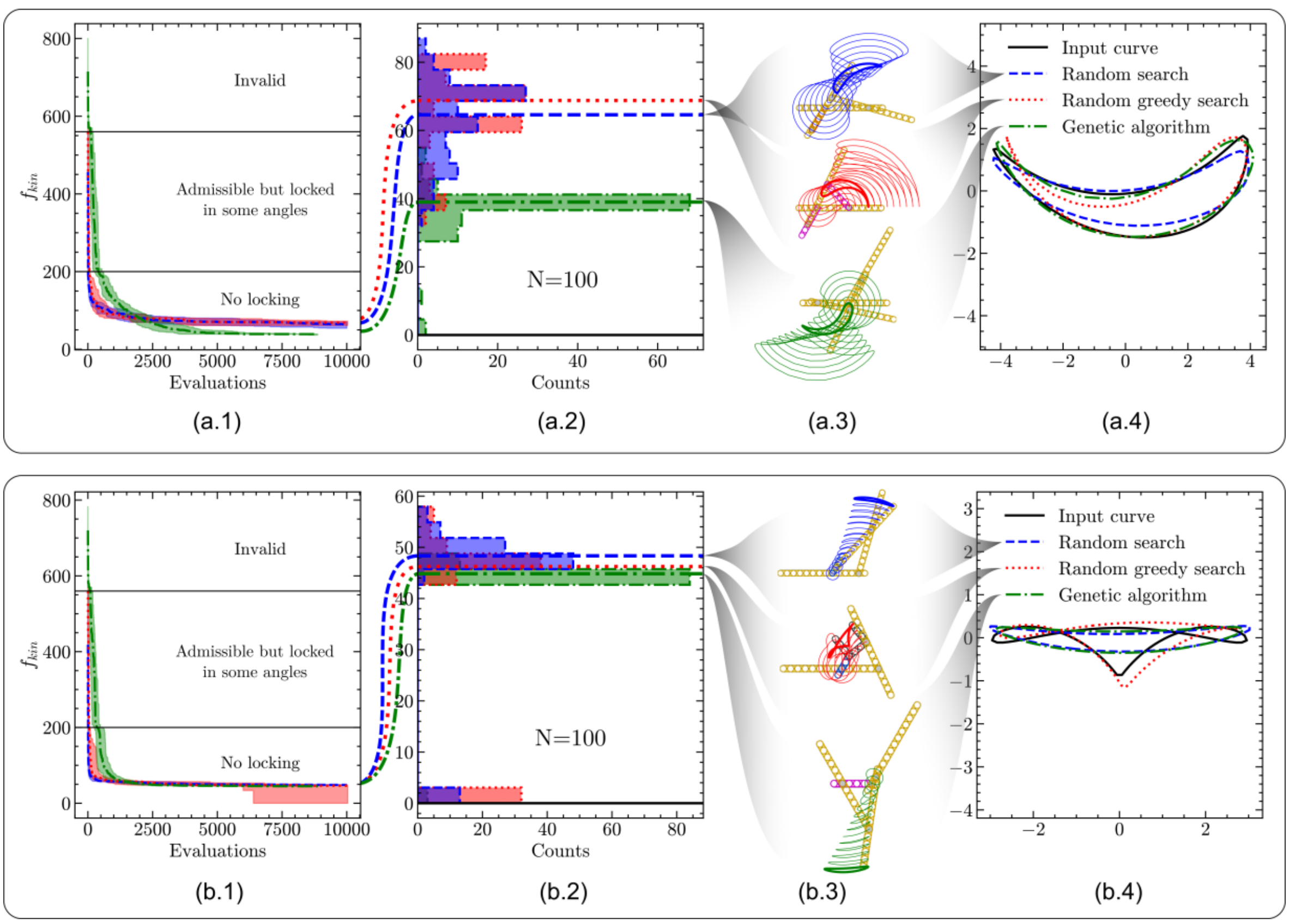

*Figure 14: (a.1-4) and (b.1-4) are the results for the first, respectively second, target curve. (a.1) and (b.1) compare the statistics of the evolution of the best $f_{kin}$ across evaluations for the three solving strategies. The medians are shown as lines while the interquartile ranges are shown as shaded areas. (a.2) and (b.2) compare the histograms of the last best $f_{kin}$ found across the 100 repetitions. (a.3) and (b.3) show the median designs. (a.4) and (a.4) compare the best matching curves of these median designs.*

Overall, it can be observed that the three solving methods provide similar results in terms of $f_{\text{kin}}$. There is no significant improvement after about 5000 evaluations, which provides an order of magnitude of when to stop the optimization. Except for the random greedy search for the second curves, all results seem to have a narrow interquartile range, meaning consistent results across the repetitions. The histograms (a.2) and (b.2) add some insight to this observation, showing that there can be rare optimization runs with a perfect result where $f_{\text{kin}}$ gets close or equal to 0, but these cases are a small minority.

Despite being imperfect matches, the median designs are good approximations of the input curve for the airfoil curve. For the second input curves, the median designs capture the scale and aspect ratio but mostly fail to respect the complexity of the curve. In both cases, the GA performs better than the random search and the random greedy search, but the superiority is marginal for the more complex, second curve.

To better understand the exploration and convergence of the three algorithms, Figure 15 displays the kinematic objective values for all designs evaluated during one run of each algorithm (10,000 evaluations), corresponding to the runs that resulted in the median designs of the first input curve. The cumulative minimum lines show what would be the best design encountered if the optimization stopped at a specific evaluation count, giving an idea of the speed of convergence. The histograms of $f_{kin}$ from the last 5,000 evaluations illustrate the distribution of performance among the designs evaluated at the final stage of the optimizations, when no significant improvements are observed.

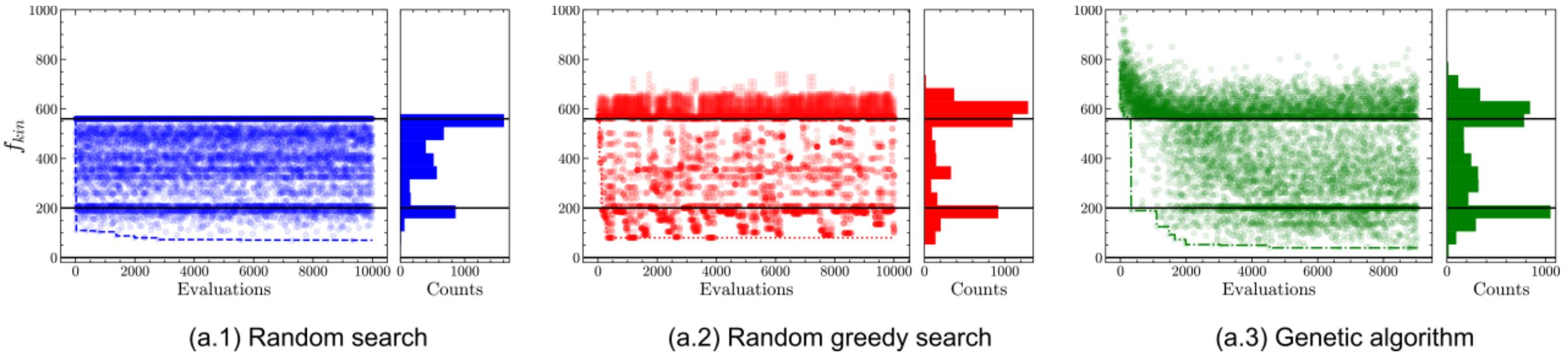

(a.1) Random search (a.2) Random greedy search (a.3) Genetic algorithm

*Figure 15: Comparing the evolution of the kinematic objective values. The dots are the $f_{kin}$ of the designs evaluated, the lines are the best $f_{kin}$ encountered until the current evaluation. For each solving strategy (a.1-3), a histogram shows the statistics of the $f_{kin}$ of the last 5,000 evaluations.*

The random search only generates admissible designs ($f_{\text{kin}} \leq 560$) because of the sampling of $\boldsymbol{p}$ and $\boldsymbol{h}$ which respects part availability, hole existence and hole availability. The histogram shows that only a fraction of the designs had no-locking ($f_{\text{kin}} \leq 360$), while the most frequent design is a locked mechanism ($f_{\text{kin}} = 560$). Even if most designs are kinematically invalid, or not good matches, the random search eventually encounters a good substitute.

The random greedy search exhibits similar behavior as the random search with few differences. Since all neighboring designs of a considered mechanism are explored, some or most of them are not even admissible. For example, changing the value of a part assigned $p_i$ might immediately result in a part that is not available or with a different hole count, incompatible with the related holes assigned. The kinematic objective can thus exceed 560 (penalty). It can also be observed that designs are grouped successively by similar values, due to the incremental improvement of the designs in the same series. More designs have no-locking than with the random search, but as observed in Figure 14(a.2), the random greedy search does not have a better performance for this target curve.

The random search is a breadth-first search with no depth while the random greedy search is depth-first search with as much breadth as authorized by the maximum evaluation. These are two different approaches resulting in different performances. In the case of this input curve, focusing on encountering as many random admissible designs as possible appears to be a better strategy than attempting to locally optimize a design, as the latter approach tends to lead to many invalid designs.

The genetic algorithm first generates mostly invalid designs but quickly reaches no-locking designs after about 1,000 evaluations and keeps improving its population until 5,000 evaluations. After 5,000 evaluations, the GA still generates a significant fraction of non-admissible designs. The explanation is like for the random greedy search. The mutation modifying a single variable is likely to render the design invalid. A crossover mixing two different designs might also create an infeasible design.

Despite generating most designs with $f_{\text{kin}} > 360$, the GA still finds better matches, as highlighted by Figure 14(a.2) and (b.2).

### 4.3. Trade-Offs

In this section, a multi-objective exploration of the kinematic performance and GHG impact of designs is performed with the GA, since it has better performance overall for the single-objective problem. The problem selected is similar as before in Section 4.2. However, this time, parts that are necessary to achieve the target curve are removed from the inventory.

Figure 16 shows the modified inventory where some parts are now unavailable. For instance, the largest parts, which are required twice to reproduce the first target curve, are now assigned a GHG value of 11.6g of $CO_2$-eq. per part used in a design.

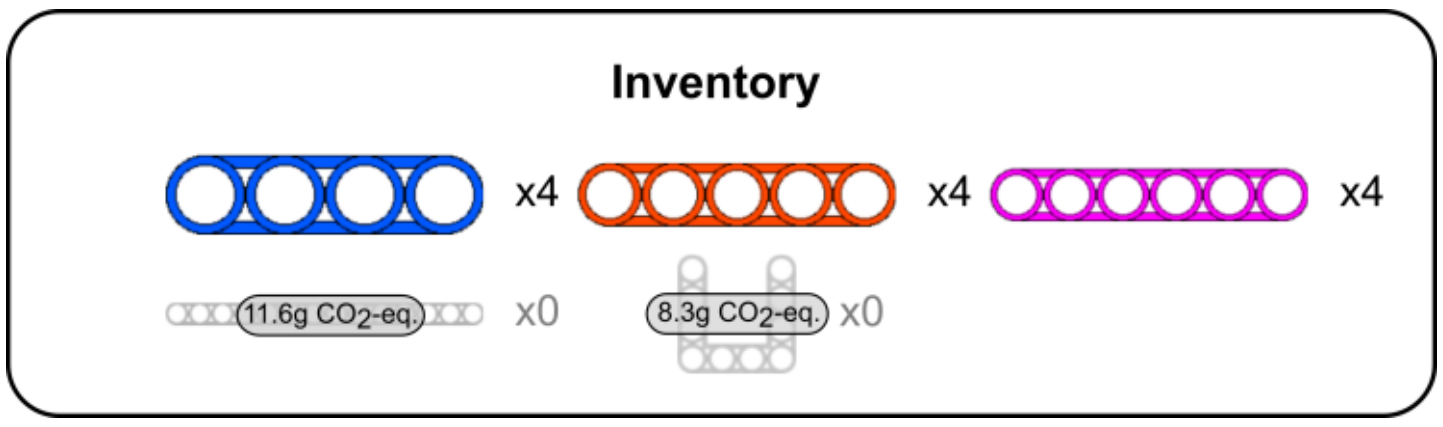

*Figure 16: Overview of the inverse design task for multi-objective optimization. The longest part and U-shaped part have been removed from the inventory and are shown in grey, i.e. they are unavailable in the stock, and have a corresponding $CO_2$ equivalent to produce them.*

The results from 400 repeated GA optimizations are presented in Figure 17. All designs across 10,000 evaluations and 400 repetitions are shown together and populate the objective space ($f_{GHG}, f_{kin}$). The NSGA-II selection strategy ensures that both objectives are explored. For comparison, all the designs that are possible to obtain by connecting the parts only at their ends are also shown. These can simply be enumerated because limiting the connections to use only the extremal holes drastically reduces the number of combinations to 750 (see Section 2.4).

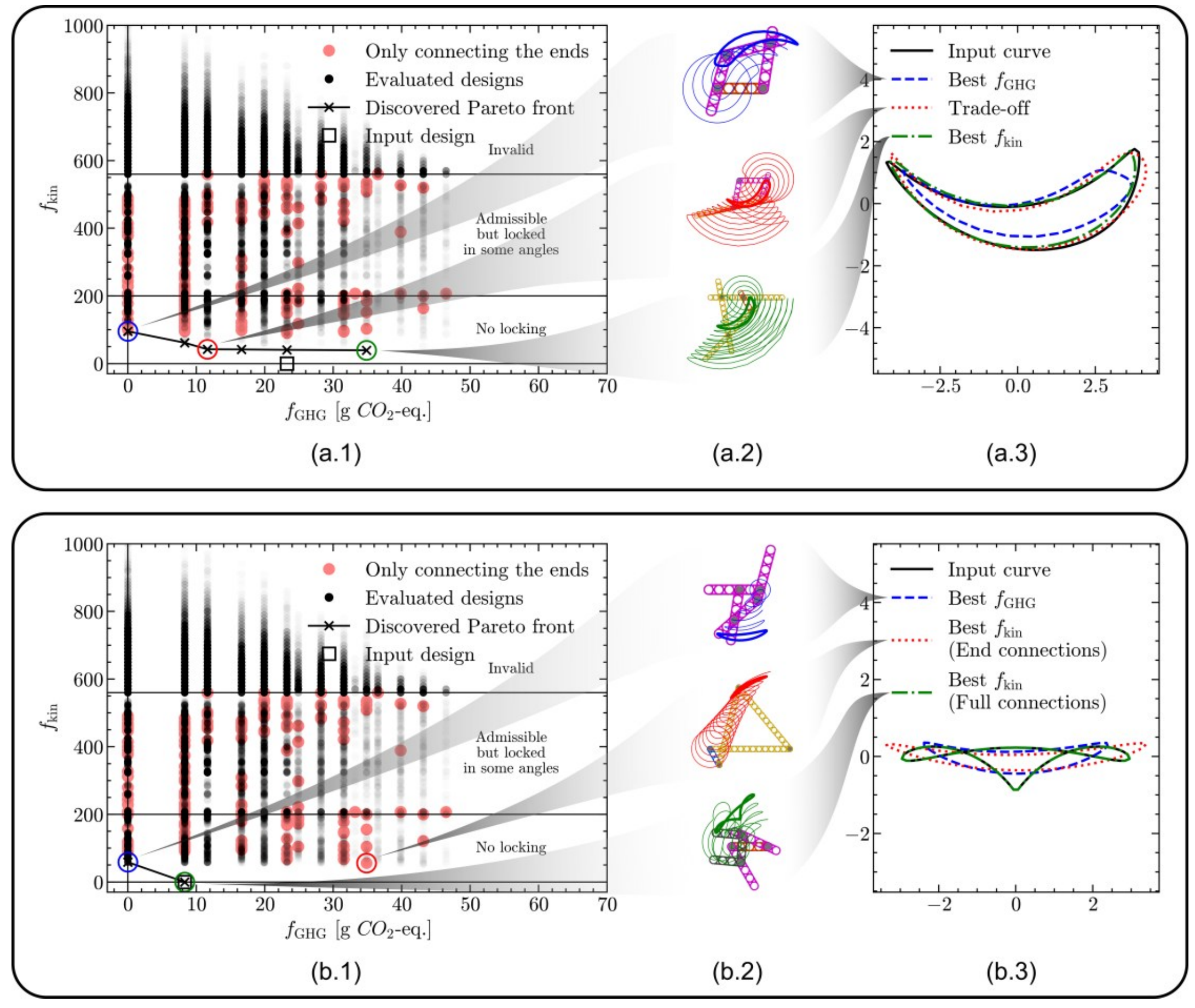

*Figure 17: Results of the multi-objective optimization for the first (a.1-3) and second (b.1-3) target curve. (a.1) and (b.1) show the objective space populated by 400 repetitions of the GA. The Pareto front of all designs evaluated is represented with the solid line. As a reference, the fully enumerated designs obtained by only connecting parts at their end holes were overlayed with bigger and lighter markers. (a.2) and (b.2) display three selected designs and (a.3) and (b.3) compare their best matching curve.*

Figure 17 (a.1) and (b.1) show designs in the objective space with quantized values for $f_{\mathrm{GHG}}$. This is because $f_{GHG}$ can only take values of multiples of the GHG impact of the unavailable parts shown in Figure 16, or a combination of them. The discovered Pareto front for the first target curve consists of six non-dominated designs, meaning that no design outperforms them in both $f_{\mathrm{kin}}$ and $f_{\mathrm{GHG}}$. However, four of these designs have a very similar $f_{\mathrm{kin}}$ value. For the second target curve, the discovered Pareto front includes only two designs: one that exclusively uses reused parts but yields a less accurate match, and another that incorporates an unavailable part to achieve a perfect match.

Among the Pareto-optimal designs for the first target curve, three designs are selected and compared: the best design using only available parts ($f_{\mathrm{GHG}} = 0$), referred to as the "best $f_{\mathrm{GHG}}$"; the design with the lowest $f_{\mathrm{kin}}$, labeled as the "best $f_{\mathrm{kin}}$", and a "trade-off" design balancing both objective. For the second target curve, the third selected design is the design with the lowest $f_{\mathrm{kin}}$ when connecting the parts only at the ends, labeled "End connections". The selected designs are shown in Figure 17(a.2) and (b.2), respectively.

Focusing on the first target curve, the comparison of the couple curves in Figure 17(a.3) shows that all three designs are a good approximation of the input curve. However, the best GHG curve is smaller than the other curves, which can be explained by the absence of the longer, required part. Although the "best $f_{\mathrm{kin}}$" design has three times the GHG impact of the "trade-off" design, the improvement in matching the curve is only marginal. This suggests that for this target curve, the "trade-off" design is already sufficient. In fact, it achieves an even lower GHG impact than the input design while still matching closely the input curve.

It is also worth noting that the input design used for the target curve could have been found, but none of the 400 GA runs discovered it. This might be explained by the fact that NSGA-II tries to satisfy both objectives and does not focus exclusively on $f_{\mathrm{kin}}$. A design with a perfect kinematic match, $f_{\mathrm{kin}} = 0$, was already rarely observed in the 100 single-objective GA, as shown in Figure 14(a.2). It can be concluded that the Pareto front shown here is not definitive, as it may exclude other dominant designs that the optimizer failed to discover.

The feasible solution space and Pareto front illustrate the inherent trade-offs between kinematic performance and environmental impact. When certain parts are unavailable in the inventory, incorporating new parts improves performance but at the expense of higher GHG emissions. This highlights the fundamental tension between optimizing functionality and minimizing environmental impact, emphasizing the need to carefully balance these competing objectives in circular design strategies.

Finally, Figure 17(a.1) and (b.1) show that all the designs with parts connected only at their ends are dominated by the designs that allow connections to intermediate holes. This suggests that the discrete flexibility of the parts is key to improving the performance of the design achievable with a certain inventory.

### 4.4. Discussion

This section discusses the results, specifically the complexity of finding a design that matches a specific curve, as well as the general implications of the results for computational circular design. The following four main challenges are identified:

- **Enumeration**: As mentioned in Section 2.4, a four-bar linkage with an inventory of five parts with a maximum of 15 holes would require more than 1.5 trillion combinations to enumerate all designs. This estimation is only for one dyad added. Even if a smarter enumeration or exploration can reduce the number of evaluations required, considering more complex topologies, bigger inventories, or more complex parts would exponentially increase this number. Restricting connections to the ends of parts makes it feasible to fully enumerate all combinations, but it also narrows the design space, limiting the variety of mechanisms that can be discovered and ultimately reducing kinematic performance.
- **Scarcity**: Most of the combinations of the design vector are invalid. Even valid designs in terms of part and hole assignments constraints, i.e. admissible designs, can be invalid kinematically. Figure 18 shows the statistics of $f_{\mathrm{kin}}$ of 100,000 randomly generated designs with no validity enforcement, unlike the random search which generates only admissible designs ($f_{kin} \leq 560$). Only 0.13% of the design space consists of mechanisms with a full 360° operating range, which demonstrates the scarcity of the design space. Although invalid designs are much quicker to evaluate, since they bypass the kinematic solver, their high prevalence in the design space means they still account for most of the computation time. To completely avoid invalid designs, the mutation and crossover operators of the GA could be adapted to preserve design admissibility, assuming the initial population is admissible.

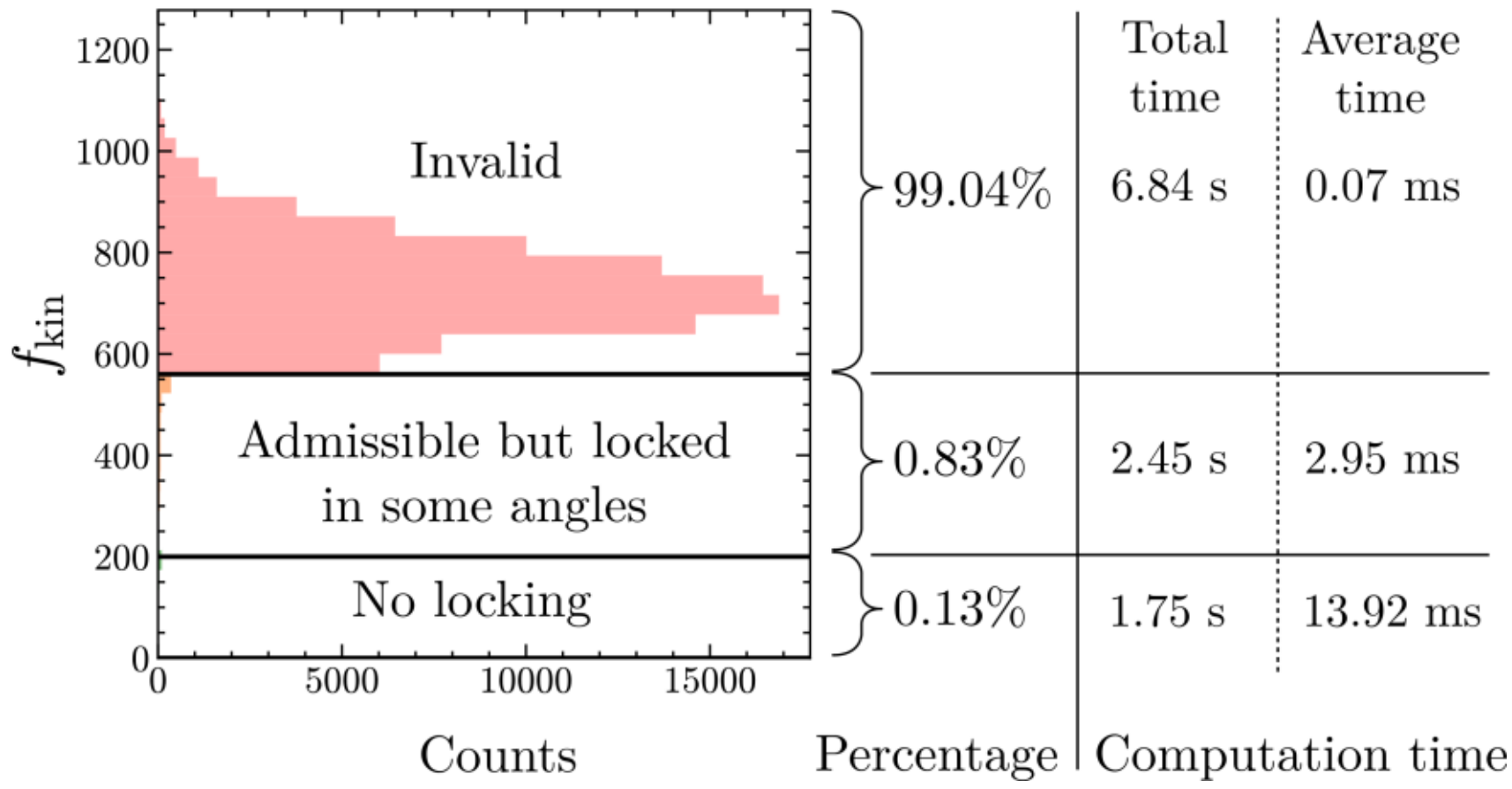

*Figure 18: Histogram of the kinematic objective values of 100,000 randomly sampled designs and associated computation times.*

- **Sensitivity**: The design space is highly sensitive to variations in design variables. As noted in Section 4.2, even a small variation in a part or hole assignment variable can lead to a significant change in the objective value. Changing a part can significantly alter the kinematics or even render the design invalid, such as when a part is too small or too large to close the mechanism. The same issue arises with changes to hole assignment variables. This results in a discontinuous design landscape beyond the discontinuous nature of integer variables.

- **Redundancy**: The same functional dimensions of a part, i.e. the distances between used holes on a part, can be achieved with different connections. This implies that a target curve can be equivalently approached with many combinations.

The inverse design problem is thus finding an optimal solution amidst a multitude of redundant, scarce local minimums in a highly variable and irregular design landscape, all within a formulation with combinatorial complexity. Even though the genetic algorithm provides candidate designs that are slightly better than the random searches, these complexities are to be addressed for more complex, real designs. This class of optimization problem, i.e. black-box, non-linear, with integer variables, is intrinsically hard to solve and seems to require here an extensive exploration to cover enough regions of the design space.

In a more realistic scenario with more complex and varied parts, the ReLink framework may require a new formulation and solving strategy. Future prospects could include using Artificial Neural Networks (ANNs) to learn the structure of the design space, as well as combining optimization and generative AI to produce designs that match the user's input.

ReLink offers designers a valuable choice between competing trade-offs. Instead of presenting a single "optimal" solution, it exposes a spectrum of alternatives that balance kinematic accuracy against sustainable, circular design. This approach is similar to travel platforms like Google Flights, which display multiple flight options, each of which highlights different trade-offs between travel time, cost, and carbon footprint, which empowers consumers to make environmentally informed choices (Crosby et al., 2024). By visualizing these alternatives, the ReLink framework helps designers identify viable options for reusing high-value components, even if it involves a slight loss in performance. This brings sustainability into the heart of the design process, encouraging more environmentally responsible practices.

## 5. Conclusion

This study presents a computational method for automatically generating, optimizing and selecting feasible designs composed of reused parts by predefining a set of standard components, exemplified by LEGO® Technic parts, and their available quantities. A tailored design representation is developed to generate mechanisms and coupler curves achievable with a given inventory of parts to reuse. To solve the inverse design problem of finding mechanisms with a coupler curve that closely matches a target curve, black-box optimization techniques are applied. Additionally, the study examines trade-offs between kinematic performance and $CO_2$ footprint when allowing new parts in the inventory.

The key challenges identified in this work include the combinatorial nature of the problem, its non-linearity, and the complexity of enforcing designs validity. By introducing an adequate formulation and solving strategies to address these challenges, this study lays the foundation for tackling more complex and realistic scenarios within the ReLink framework. More broadly, the proposed framework paves the way for a new area of research in the computational circular design of mechanical products and provides a new benchmark problem. It marks a significant step toward developing a comprehensive framework for computational circular design that is envisioned to support design practitioners in creating mechanical systems that integrate reused components, contributing to cleaner and more sustainable production practices.

## Acknowledgements

The authors would like to thank Marc Wirth and Rafaela Louis for proofreading the manuscript and helping improve its clarity through their valuable suggestions.

## CRediT authorship contribution statement

**Maxime Escande**: Writing - original draft, Visualization, Software, Validation, Investigation, Conceptualization, Methodology, Data curation, Formal analysis. **Kristina Shea**: Writing - review & editing, Supervision, Methodology, Conceptualization, Project administration.

## Data Availability

The Python source code and implementation framework developed for the methods introduced in this paper are openly available in the GitHub repository: *https://github.com/ETHZ-EDAC/ReLink* and deposited on Zenodo (Escande and Shea, 2026).

## Funding

This research did not receive any specific grant from funding agencies in the public, commercial, or not-for-profit sectors.

## Declaration of Competing Interest

The authors declare that they have no known competing financial interests or personal relationships that could have appeared to influence the work reported in this paper.

## Declaration of use of AI in the writing process.

During the preparation of this work the authors used ChatGPT (OpenAI) to improve the clarity and readability of the text. After using this tool, the authors reviewed and edited the content as needed and take full responsibility for the content of the published article.

# Supplementary Information

## A.1. Normalization procedure

During *trajectory resampling*, the trajectories are resampled to $\hat{n}$ uniformly spaced discrete curve points. During *position normalization*, the curve is shifted such that the center of gravity is at the origin. During o*rientation normalization*, the curve is rotated such that the principal axis of lower inertia is aligned along the horizontal coordinate axis. The moments of inertia are represented according to Equation $A.9$.

$$\mathbf{I} = \begin{bmatrix} a & b \\ b & c \end{bmatrix}, \text{with } a = \sum_{i=0}^{n-1} x_i^2, b = \sum_{i=0}^{n-1} x_i\, y_i, \text{and } c = \sum_{i=0}^{n-1} y_i^2 \qquad (A.9)$$

Using $D = \sqrt{4b^2 + (a - c)^2}$, the angle of the principal axis associated with the lower eigenvalue, or minimum inertia, is computed according to Equation $A.10$.

$$\alpha = \text{atan2}(2b, a - c - D) \qquad (A.10)$$

The curve can then be aligned along $\alpha$ or $\alpha + \pi$.

During *direction normalization*, the curve is either either maintained in its original orientation or rotated by 180° to ensure that its central top point is positioned as low as possible. This ensures consistent curve directionality, following the method described in Nobari et al. (2022).

## A.2. Penalty terms and weights

The weights of the objective function in Equation 4 are $w_{\text{CD}} = 200$, $w_1 = w_2 = w_3 = w_4 = 10$.

Table 2 lists the penalty terms and their meaning.

*Table 2: List of the penalty terms*

| Symbol | Penalty name | Description |
|---|---|---|
| $P_1(\boldsymbol{p})$ | Part availability | The part availability penalty ensures that the parts assigned are available in the inventory. This penalty term quantifies the number of parts that exceeds the available counts of each part type in the inventory. |
| $P_2(\boldsymbol{h})$ | Hole availability | The hole availability penalty ensures that each hole in a part is assigned to at most one pin connector by counting the number of pin connectors redundantly attached to the same part hole. If two pin connectors were assigned to the same hole, the part concerned might rotate independently from the rest of the mechanism. |
| $P_3(\boldsymbol{p}, \boldsymbol{h})$ | Hole existence | The hole existence penalty ensures that a part does not reference a hole index exceeding its hole count – for instance, a part with four holes cannot have a connection assigned to an inexistent fifth hole. It sums all indices overshoots across all part vertices. |
| $P_4(\boldsymbol{p}, \boldsymbol{h})$ | Geometric validity | The geometric validity penalty penalizes the mechanisms that experience locking, meaning their operating range is not complete. The operating range is evaluated against the full range $(0°, 360°)$: $$P_4 = 360 - (\theta_{max} - \theta_{min})$$ where $(\theta_{min}, \theta_{max})$ is the operating range, expressed in degrees. |

## A.3. Parameters of the Genetic Algorithm

Table 3 lists the parameters of the genetic algorithm used in Section 4.2.

*Table 3: Parameters of the genetic algorithm*

| Symbol | Value | Description |
|---|---|---|
| $N_{\text{pop}}$ | 200 | Population size |
| $N_{\text{elit}}$ | 3 | Number of best designs kept permanently |
| $p_{\text{mut}}$ | 10% | Probability of mutation (random mutation) |
| $p_{\text{cros}}$ | 60% | Probability of crossover (single-point crossover) |
| $N_{\text{mating}}$ | 100 | Number of designs to be selected as parents (steady-state selection for single-objective, NSGA-II for multi-objective) |